\pdfoutput=1
\documentclass[
superscriptaddress,
twocolumn,
showpacs,preprintnumbers,
amsmath,amssymb,
aps,
prl,
]{revtex4-2}
\usepackage{graphicx}% Include figure files
\usepackage{dcolumn}% Align table columns on decimal point
\usepackage{bm}% bold math

\usepackage{float}

\begin{document}

%\preprint{APS/123-QED}

\title{2D hybrid CrCl$_2$(N$_2$C$_4$H$_4$)$_2$ with tunable ferromagnetic half-metallicity}

% Force line breaks with \\

\author{Wentao Hu}
 \affiliation{Laboratory for Computational Physical Sciences (MOE),
 State Key Laboratory of Surface Physics, and Department of Physics,
 Fudan University, Shanghai 200433, China}

\author{Ke Yang}
\affiliation{College of Science, University of Shanghai for Science and Technology, Shanghai 200093, China}
 \affiliation{Laboratory for Computational Physical Sciences (MOE),
 State Key Laboratory of Surface Physics, and Department of Physics,
  Fudan University, Shanghai 200433, China}

\author{Alessandro Stroppa}
\email{Corresponding author. alessandro.stroppa@aquila.infn.it}
 \affiliation{Consiglio Nazionale delle Ricercheâ CNR-SPIN, 67100 - Coppito (L'Aquila), Italy.}
 \affiliation{Department of Physical and Chemical Sciences, Universit\`a degli Studi dell'Aquila, 67100 - Coppito (L'Aquila), Italy}

\author{Alessandra Continenza}
\affiliation{Department of Physical and Chemical Sciences, Universit\`a degli Studi dell'Aquila, 67100 - Coppito (L'Aquila), Italy}

\author{Hua Wu}
\email{Corresponding author. wuh@fudan.edu.cn}
\affiliation{Laboratory for Computational Physical Sciences (MOE),
 State Key Laboratory of Surface Physics, and Department of Physics,
 Fudan University, Shanghai 200433, China}
\affiliation{Collaborative Innovation Center of Advanced Microstructures,
 Nanjing 210093, China}

%\affiliation{
%\affiliation{
%\affiliation{%

\date{\today}

\begin{abstract}

Two-dimensional ferromagnetic (2D FM) half-metal holds great potential for quantum magnetoelectronics and spintronic devices.
Here, using density functional calculations and magnetic pictures, we study the electronic structure and magnetic properties of the novel van der Waals (vdW) metal-organic framework (MOF), CrCl$_2$(N$_2$C$_4$H$_4$)$_2$, \textit{i.e.} CrCl$_2$(pyrazine)$_2$.
Our results show that CrCl$_2$(pyrazine)$_2$ is a 2D FM half-metal, having a strong intralayer FM coupling but a much weak interlayer one due to the vdW spacing. Its spin-polarized conduction bands are formed by the pyrazine molecular orbitals and are polarized by the robust Cr$^{3+}$ local spin = 3/2. These results agree with
the recent experiments [Pedersen \textit{et al.}, \textit{Nature Chemistry}, 2018, \textbf{10}, 1056].
More interestingly, CrCl$_2$(pyrazine)$_2$ monolayer has a strong doping tunability of the FM half-metallicity, and the FM coupling would be significantly enhanced by electron doping.
Our work highlights a vital role of the organic ligand and suggests that vdW MOF is also worth exploration for new 2D magnetic materials.
\end{abstract}

%\pacs{}% PACS, the Physics and Astronomy
\maketitle

%\tableofcontents

%\section{I. Introduction}
%%introduction

\section{Introduction}
The possibility to achieve manipulation of magnetic properties through changes of the structure of materials has always been an attractive topic for basic and applied research in material science.
In particular, transition-metal based inorganic compounds offer a wide playground where electronic and magnetic properties could be tuned to achieve novel phenomena such as superconductivity\cite{wang2018evidence}, quantum Hall effect\cite{wu2018observation}, topological insulators\cite{tokura2019magnetic} and multiferroicity\cite{spaldin2019advances}.
The multifunctional properties are often linked to the interplay of charge, orbital and spin degrees of freedom.\cite{khomskii2014transition}
Therefore, transition-metal atoms represent essential ingredients of several technologically interesting materials\cite{vsmejkal2018topological,sun2020self,manchon2019current}.
Hybrid compounds, \textit{i.e.,} compounds showing coexistence of organic and inorganic components, further increase the possibility to tune physical properties of materials, thus enlarging the horizons for possible device applications.
For example, yet another interesting approach to tune structural, electronic and magnetic properties is to explore the effects of the ligands.
Recently, a promising class of materials has emerged such as metal-organic frameworks (MOFs)\cite{Li2016MOFreview,yuan2018stable,coronado2020molecular,Stroppa2014ferroelectric}. They are made up of a network of metal ions bridged by organic ligands, forming a porous framework.
In these materials, different ligands can lead to totally different conducting and/or magnetic properties.
Eventually, the organic ligands retains a free-radical character, thus making the hybrid compound conductive\cite{darago2015electronic,ma2019using}.

After the successful exfoliation of graphene, two-dimensional materials have become one of the hottest research field in the last decades, due to their novel and diverse physical properties\cite{li2017direct,huang2017layer,gong2017discovery,deng2018gate}.
In particular, ferromagnetism has been recently observed in several new layered inorganic materials, such as monolayer CrI$_3$\cite{huang2017layer} and few layer Cr$_2$Ge$_2$Te$_6$\cite{gong2017discovery}, both of them showing wide application potential.
Efforts on exfoliating similar materials were reported recently\cite{fei2018Fe3GeTe2,Lee2016FePS3,Lin2016CrSiTe3,Bonilla2018VSe2,kazim2020mechanical,serri2020enhancement}.
Also, theoretical studies try to understand, predict and utilize the 2D magnetic properties\cite{Lado2017anisotropy,Kim2019LS,Jiang2019stacking,Sivadas2018stacking,Chengxi2018Toward,Zhuang2016anisotropy,yang2020VI3,liu2020VBr3}.
Therefore, 2D magnetic materials are still a rapid growing and developing field\cite{burch2018review,Cheng2019review}.

\begin{figure}[t]
	\includegraphics[width=8.6cm]{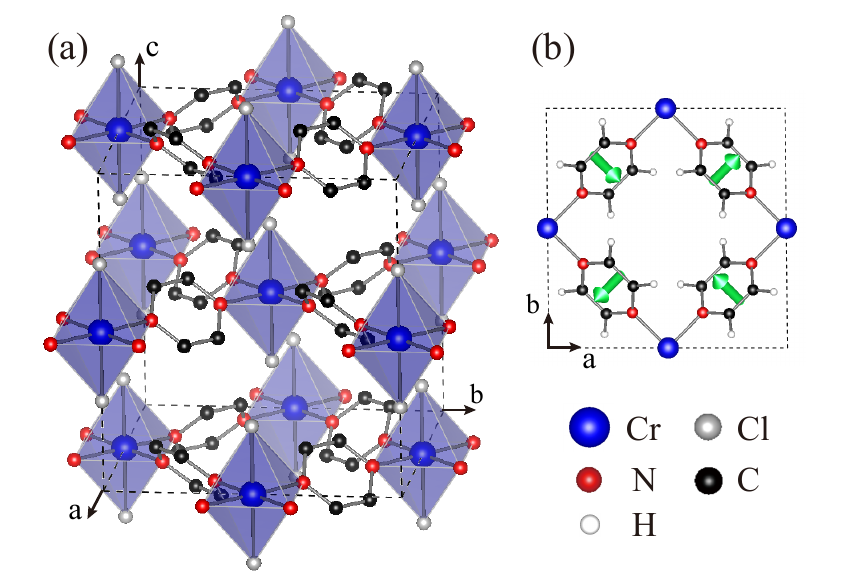}
	\caption{(a) Side view of the bulk CrCl$_2$(pyrazine)$_2$. (b) Top view into the $ab$ plane of the top layer. Hydrogen atoms in (a) and Chlorine atoms in (b) are hidden for simplicity.}{\label{bulk_structures}}
\end{figure}

Very recently, hybrid materials have joined to the 2D materials landscape\cite{Zhao2018MOF2D,Luo2019HierarchicalMOF,pedersen2018formation,perlepe2020}.
This certainly defines new directions to explore, since the dual organic-inorganic nature of the materials together with the dimensionality decrease from 3D to 2D adds new functional and structural flexibility as well as increases the tunability of relevant physical properties.
Therefore, it is of great interest to search for new 2D materials starting from bulk layered materials which could be easily exfoliated into monolayer.
Recently, a new bulk but layered material CrCl$_2$(pyrazine)$_2$ has been synthesized\cite{pedersen2018formation}.
This compound is very promising because it not only shows magnetic properties related to both transition metal and organic ligands, but also could be exfoliated into a new 2D hybrid material.
According to experimental measurements\cite{pedersen2018formation}, CrCl$_2$(pyrazine)$_2$ is a ferromagnetic (FM) metal with Curie temperature $T_{\rm C}$ $\simeq$ 55 K.
All these considerations suggest that this new material is suitable for nanotechnology applications and is therefore calling for a deeper theoretical study.
This represents the main motivation of our work.

In this article, we provide new insights on the structural, electronic and magnetic properties of bulk and monolayer of CrCl$_2$(pyrazine)$_2$ using density functional theory (DFT) calculations.
Our results show that the bulk is a robust half-metal with strong intralayer FM and weak interlayer coupling, which come from molecular orbitals induced by a hybridization between Cr-$3d$ and pyrazine molecular orbitals.
More interestingly, the FM coupling of CrCl$_2$(pyrazine)$_2$ monolayer can be significantly enhanced by electron doping, but can also be changed into antiferromagnetic (AF) state by hole doping.
Therefore, we predict that CrCl$_2$(pyrazine)$_2$ would be an appealing 2D spintronic material.

\section{Computational details}
%{\label{Computational details}}
Density functional theory (DFT) calculations were carried out using the Vienna Ab-initio Simulation Package (VASP)\cite{kresse1996}.
The wave function was expressed with the plane-wave basis set and a cut-off energy of 450 eV was used.
The exchange and correlation energy was described by the generalized gradient approximation (GGA) with the Perdew, Burke, and Ernzerhof functional\cite{PBE}.
To better describe the on-site Coulomb interactions of Cr 3$d$ electrons, the typical value of the Hubbard $U = 4.0$ eV and Hund exchange $J = 0.9$ eV were used in the GGA+U calculations\cite{U}.
$\sqrt{2} \times \sqrt{2} \times$ 1 supercell was chosen for bulk in order to study different magnetic structures.
For monolayer, a lateral $\sqrt{2}\times\sqrt{2}$ supercell was chosen with a vacuum of 7 {\AA}.
The Monkhorst-Pack k-mesh of 5${\times5\times4}$ (5${\times5\times1}$) was used for bulk (monolayer) calculations.
The total energy converged to 10$^{-5}$ eV and all the atoms were fully relaxed till the forces converged to 0.01 eV/\AA.
The PBE-D2 corrections within the Grimme's approach was used for the cleavage energy calculation\cite{VdW-D2}.

%%------------------------------------%%
\section{Results and discussions}
\subsection{CrCl$_2$(pyrazine)$_2$ bulk}

\begin{table}[b]
	\caption{Relative total energies ${\Delta \emph E}$ (meV/f.u.), total spin moments (${\rm \mu_B/f.u.}$) and local spin moments (${\rm \mu_B}$) for the bulk structure in the FM or intralayer-AF state given by GGA+U calculations. The organic ligands are polarized by Cr$^{3+}$ spin = 3/2, giving a small opposite local spin moment at each N atom. The marginal C and H moments are not shown.}
	\label{table_bulk}
	\begin{tabular*}{0.48\textwidth}{@{\extracolsep{\fill}}crrrr}
		\hline
		states&${\Delta \emph E}$&total&Cr&N \\
		\hline
		FM & 0 & 2.00 & 2.69 & --0.10  \\
		AF & 163 & 0.00 & $\pm$2.68 & $\mp$0.02  \\
		\hline
	\end{tabular*}
\end{table}

%GGA
We start with the bulk CrCl$_2$(pyrazine)$_2$ for which the experimental results are available for comparison\cite{pedersen2018formation}.
The bulk is a van de Waals (vdW) material with AB stacking, see Fig.\ \ref{bulk_structures}.
The Cr ion is surrounded by four pyrazines in a,b plane and two Cl ions along c axis.
It has a local distorted octahedra, which splits the Cr-3$d$ orbitals into $t_{2g}$ triplet and $e_g$ doublet.
We first perform  GGA calculations with spin-polarization.
We carry out a full structural optimization for bulk CrCl$_2$(pyrazine)$_2$ in four different structures, see Table S1 in Supporting Information (SI). Our results show that $\alpha$ structure is most stable, which is the case in the monolayer as detailed in section 3.2. The optimized lattice constants agree well with the experimental ones.
The Cr local spin moment is 2.35 ${\rm \mu_B}$ which is reduced from Cr$^{3+}$ $S = 3/2$ state by a covalence.
N (C) local spin moment is --0.08 ${\rm \mu_B}$ (--0.01 ${\rm \mu_B}$) which is polarized by the Cr spin.
It is important to note that the total spin moment is 2.00 ${\rm \mu_B}$ per formula unit (f.u.) which is indicative of an antiparallel ${S}$ = --1/2 contribution from the organic ligands.
The total magnetic moment agrees with the experimental one of 1.8 ${\rm \mu_B}$\cite{pedersen2018formation}.
Note that the magnetic ground state of CrCl$_2$(pyrazine)$_2$ is ferrimagnetic\cite{pedersen2018formation}, with opposite spins of Cr and pyrazines. However, to better describe the effective Cr-Cr FM coupling and compare it with a possible Cr-Cr AF state, we refer to the ferrimagnetic ground state as the FM state throughout the main text.

%+U
To better describe the correlated Cr 3$d$ electrons, we perform GGA+U calculations.
The local Cr$^{3+}$ spin moment is now increased up to 2.69 ${\rm \mu_B}$, see Table\ \ref{table_bulk}.
Again, the total spin moment is 2.00 ${\rm \mu_B}$/f.u., which well corresponds to the Cr$^{3+}$ $S = 3/2$ and the induced opposite ${S}$= --1/2 on the organic ligands.
In order to estimate the magnetic coupling in the bulk CrCl$_2$(pyrazine)$_2$, we calculated FM, interlayer-AF (with intralayer-FM) and intralayer-AF state by GGA+U.
We find that FM is the ground state which accords with the experiment\cite{pedersen2018formation}.
The intralayer-AF state turns out to be much less stable than the FM state by 163 meV/f.u., demonstrating a strong intralayer FM coupling.
In contrast, the interlayer coupling is much weak due to the vdW spacing, with the interlayer-AF being 6 meV/f.u. higher than the FM ground state.

We also check the spin-orbit coupling (SOC) effect. The GGA+U and GGA+U+SOC results are practically the same, see the band structures in Fig. S1 in SI. In addition, the interlayer-AF state is less stable than the FM ground state by 163 (137) meV/f.u. for bulk (monolayer) by GGA+U+SOC, which is (almost) the same as the GGA+U results of 163 (136) meV/f.u. This is due to the negligible SOC effects of the closed Cr$^{3+}$ $t_{2g}^3$ shell and the pyrazine molecule with the light C/N/H atoms.

\begin{figure}[t]
	\includegraphics[width=8.6cm]{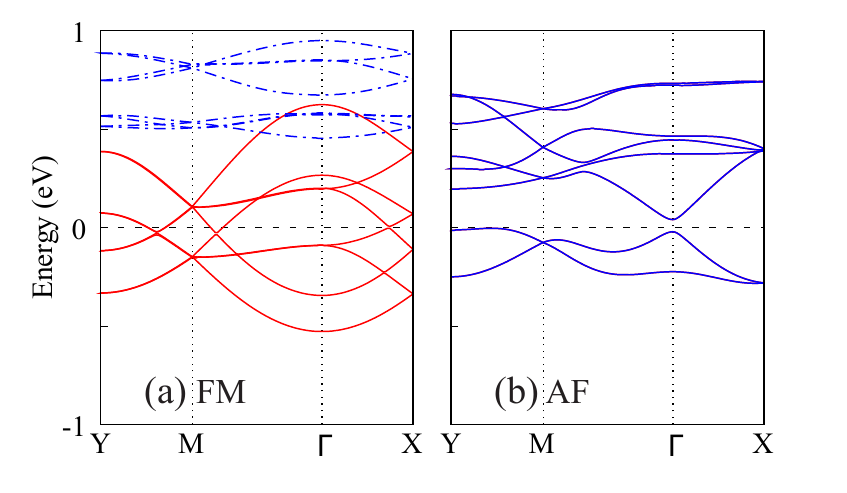}
	\caption{Band structures of (a) the FM state and (b) the intralayer-AF state in A-type bulk CrCl$_2$(pyrazine)$_2$, calculated by GGA+U. The blue (red) lines stand for the up (down) spin channel. The Fermi level is set at zero.}{\label{bulk}}
\end{figure}

\begin{figure}[t]
	\includegraphics[width=8.6cm]{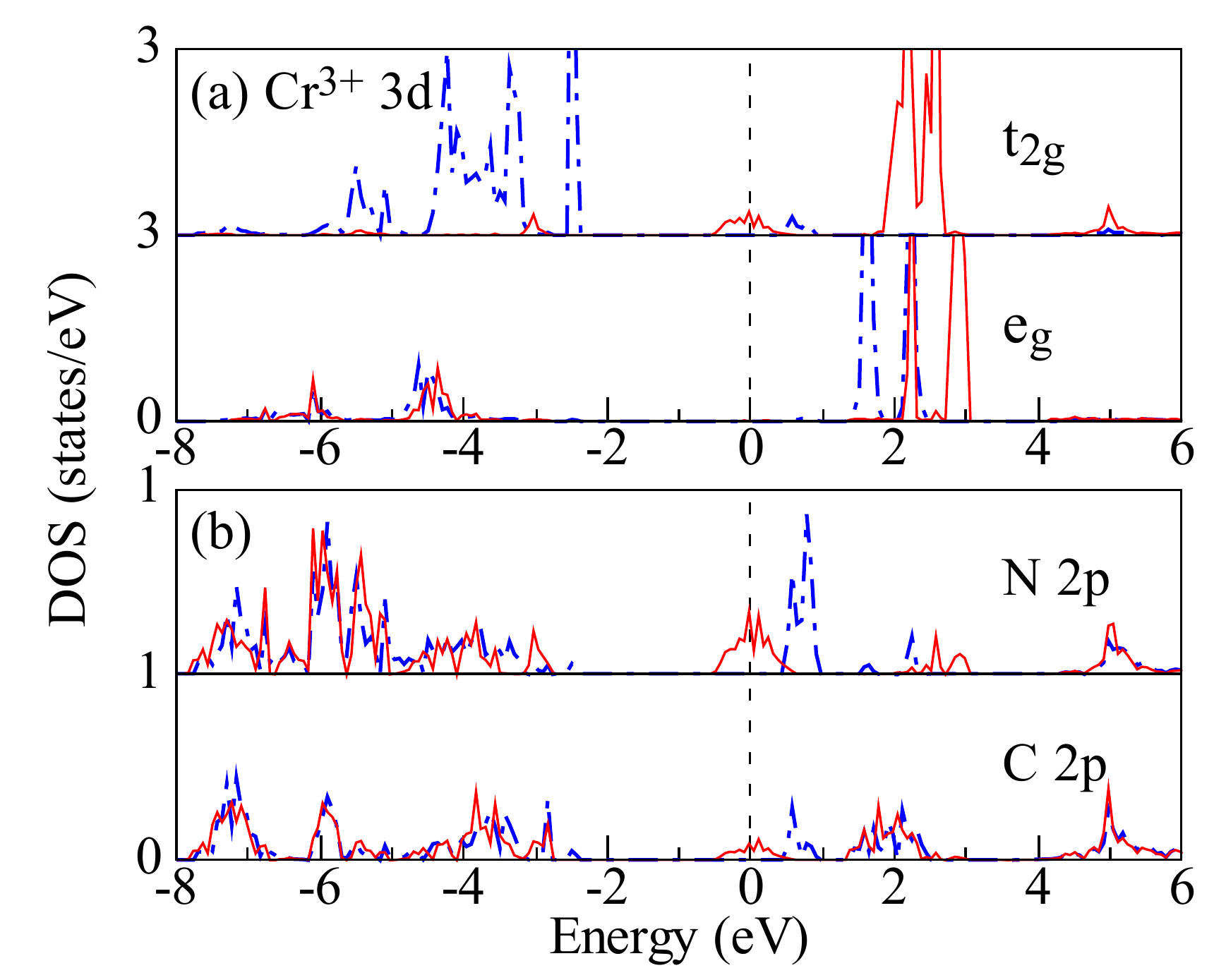}
	\caption{(a) Cr$^{3+}$ 3$d$ (b) N and C 2$p$ density of states (DOS) calculated by GGA+U. The blue (red) lines stand for the up (down) spin. The Fermi level is set at zero. }{\label{pDOS-bulk}}
\end{figure}

%FM half-metal
Next, to study the electronic properties of the material, we plot the band structure for the FM and intralayer-AF state, see Fig.\ \ref{bulk}.
A clear half metal is demonstrated in the FM state since eight down-spin bands crossing the Fermi level, and a large up-spin band gap of more than 3 eV can be observed, see also Fig.\ \ref{pDOS-bulk}.
This coincides with the reported high electronic conductivity in experiment\cite{pedersen2018formation}.
Due to this electronic itinerancy, the down-spin bands show more dispersion compared with the up-spin ones.
Note that owing to the similarity between layers, all these bands show similar curves in pairs which are dispersed by the weak interlayer interaction.
In contrast, the intralayer-AF state has a less band dispersion and becomes an insulator with a small energy gap.
Note that we also perform the hybrid functional HSE06 calculations for a comparison with the GGA+U results, see Fig. S2 in SI. The major FM half-metallicity remains unchanged in both functionals, and the shape of the band structure crossing the Fermi level is quite similar. The calculated spin moments are close, 2.73 ${\rm \mu_B}$ vs 2.69 ${\rm \mu_B}$ for the Cr$^{3+}$ (--0.12 ${\rm \mu_B}$ vs --0.10 ${\rm \mu_B}$ for the N atom).

%pDOS
In order to further analyze the band composition near the Fermi level, we plot in Fig.\ \ref{pDOS-bulk} the orbitally resolved density of states (DOS) of the FM ground state.
For Cr 3$d$ states there is a clear splitting between $t_{2g}$ and empty $e_g$, and the half-occupied $t_{2g}$ shell confirms the high-spin configuration.
DOS intensity across the Fermi level can be found in N 2$p$ and C 2$p$ down-spin channel, which corresponds to the down-spin bands in Fig.\ \ref{bulk}(a).
Also, there is a small DOS intensity from Cr $t_{2g}$ down-spin states, which suggests a hybridization between Cr and pyrazines.
Note that hybridization states can also be found in the up-spin channel, but lie about 0.5 eV higher than the down-spin one due to an exchange splitting induced by the FM Cr sublattice.
Hence, those bands across the Fermi level, being split by Cr polarization, are dominated by N $2p$ and C $2p$ states hybridized with Cr $t_{2g}$. These results highlight the vital role of the organic ligands in the FM half-metallicity.

%%------------------------------------%%
\subsection{CrCl$_2$(pyrazine)$_2$ monolayer }

\begin{figure}[t]
	\includegraphics[width=8.6cm]{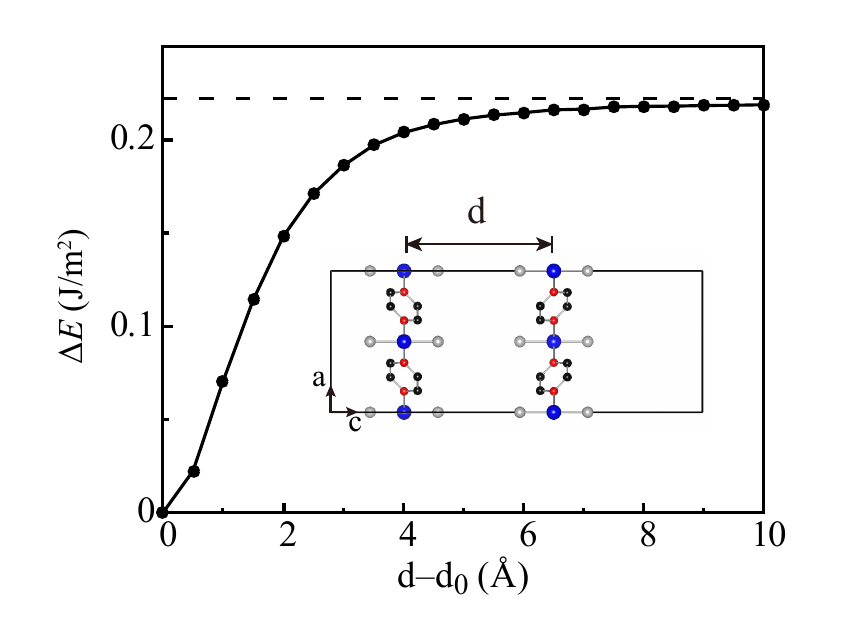}
	\caption{The relative total energy calculated as a function of the distance between two CrCl$_2$(pyrazine)$_2$ monolayers with a reference to the experimental vdW distance d$_0$. The cleavage energy is estimated to be 0.22 J/m$^2$.}
	{\label{ML_cleave}}
\end{figure}

\begin{figure}[t]
	\includegraphics[width=8.6cm]{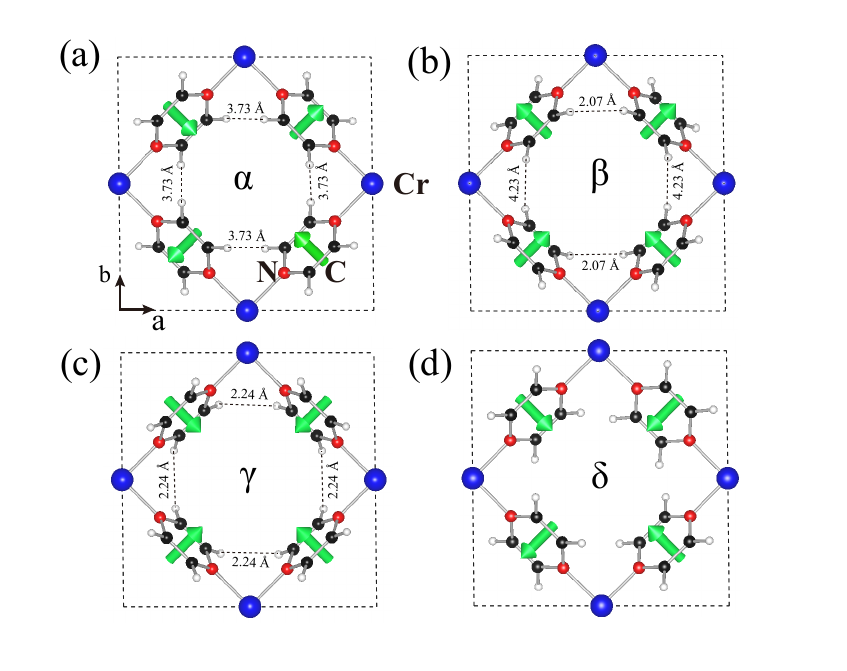}
	\caption{Top view into $ab$ plane of (a) $\alpha$-type, (b) $\beta$-type (c) $\gamma$-type and (d) $\delta$-type monolayer structures. All structures (except $\delta$) are after relaxation. Green arrows are perpendicular to the planes of the pyrazine. Chlorine atoms are hidden for simplicity.}
	{\label{ML_structures}}
\end{figure}

Motivated by the above finding of the FM half-metallicity in the vdW MOF and the strong intralayer (weak interlayer) FM coupling , we now study the CrCl$_2$(pyrazine)$_2$ monolayer which could well be an interesting 2D magnetic material. Here we calculate the cleavage energy using the PBE-D2 correction, see Fig.\ \ref{ML_cleave}.
The total energy results as a function of the increasing interlayer distance allow us to estimate the cleavage energy, and it is 0.22 J/$m^2$ and is even lower than 0.3 J/$m^2$ for CrI$_3$\cite{mcguire2015coupling} which has been successfully exfoliated from the bulk.
Therefore, an exfoliation of the CrCl$_2$(pyrazine)$_2$ is likely, and we now explore the electronic and magnetic properties of the monolayer by GGA+U calculations and then establish a physical picture.

%structure
We first investigate how the pyrazines affect the structural energy and magnetic order, and at the same time search the stable structure of the monolayer.
In CrCl$_2$(pyrazine)$_2$ each pyrazine has two possible orientations, and after taking symmetry into account there exist four different possible monolayer structures $\alpha, \beta, \gamma$ and $\delta$, see Fig.\ \ref{ML_structures}.
$\alpha$ is the layer component of our bulk structure, containing two pairs of pyrazines in different orientations.
$\gamma$ also contains two pairs of pyrazine in different orientations, but has exchanged one pair of pyrazines from $\alpha$.
$\beta$ has two pairs of pyrazines in the same orientations, and $\delta$ has one different pair of pyrazines and one same pair.
After atomic relaxations, the results of total energy calculation (see Table\ \ref{table_ML}) show that $\alpha$ is the energetically most favorable structure.
For comparison, energy of $\beta$ and $\gamma$ is respectively 182 and 302 meV against $\alpha$, while the initialized $\delta$ structure is unstable and converges to $\alpha$.
The different structural energies arise from different ligand repulsion, which is related to the H ion distance of adjacent pyrazines.
All the adjacent pyrazine pairs in $\alpha$ locally avoid each other and thus effectively lower the repulsion energy.
In contrast, instability is induced by stronger repulsion in $\beta$ and $\gamma$ since respectively two and four pairs of H ions have a much closer distance compared with $\alpha$.
FM ground states can be found for all monolayer structures, which is consistent with the bulk case.
Moreover, the most favored $\alpha$ structure also has the largest relative magnetic energy (136 meV/f.u.), indicating that a stable distribution of pyrazine orientations could benefit the FM coupling.
Notice, the FM coupling of $\alpha$ is comparable with bulk (163 meV/f.u.), since the former is the layer component of the later one.

\begin{figure}[h]
	\includegraphics[width=8.6cm]{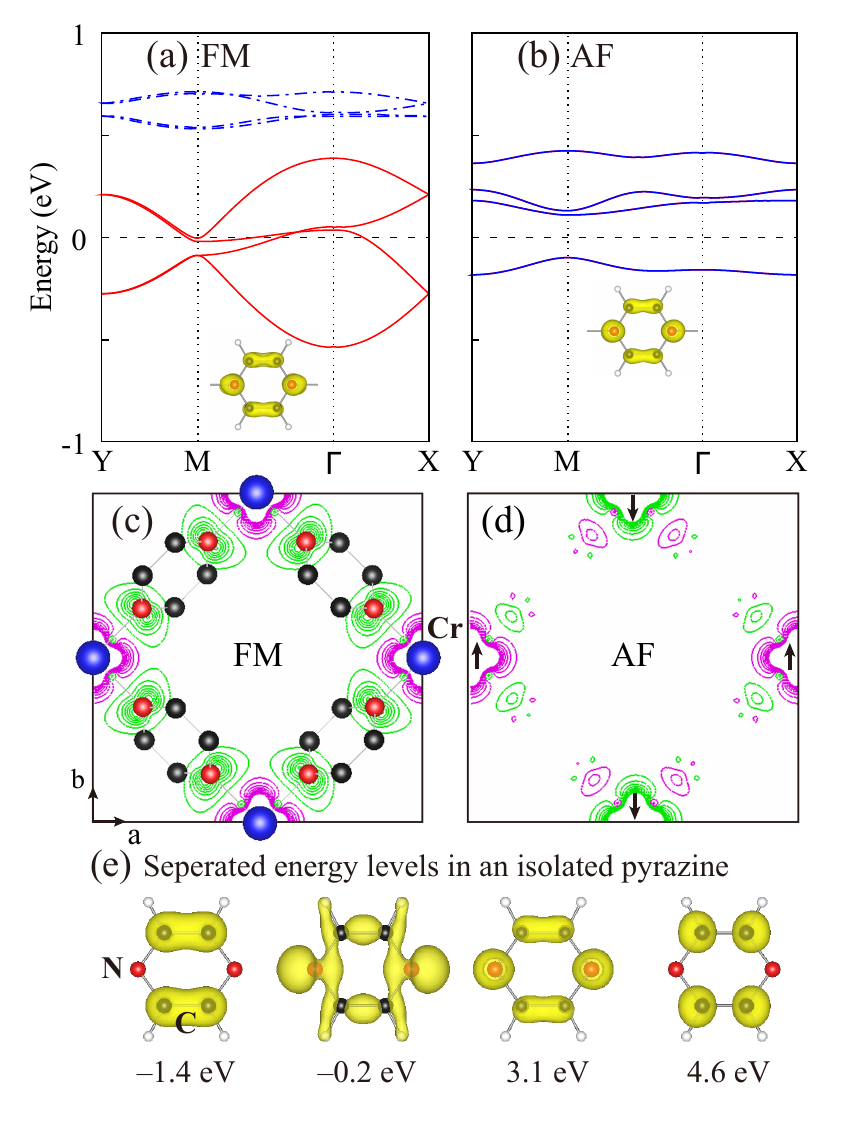}
	\caption{Band structures of (a) the FM state and (b) the AF state in $\alpha$-type monolayer CrCl$_2$(pyrazine)$_2$, calculated by GGA+U. The blue (red) lines stand for the up (down) spin. The Fermi level is set at zero. The inserts in (a) and (b) are charge density plots of the pyrazine part for one of the bands near the Fermi level. Spin density plots of (c) the FM state and (d) the AF state. The purple (green) lines are contours for up (down) spin density. The black arrows represent the spin moment directions of Cr. (e) Charge density plots and separated energy levels of the four molecular orbitals nearest to the Fermi level in an isolated molecular pyrazine, by a single k-point calculation.}{\label{monolayer}}
\end{figure}

%FM half-metal
Hereafter, we focus on the most stable $\alpha$ structure, exploring the electronic and magnetic properties of monolayer CrCl$_2$(pyrazine)$_2$.
In Fig.\ \ref{monolayer} (a) and (b) we present the band structure for the two magnetic configurations.
The FM ground state, with four down-spin bands crossing the Fermi level (and an up-spin gap of 2.5 eV, not shown), is predicted to be a robust half-metal.
On the other hand, the AF state is insulating with a much reduced bandwidth.
To clearly show the magnetic alignment, we plot in Fig.\ \ref{monolayer} (c) and (d) the spin density for the FM and AF state.
As expected for a high-spin $S = 3/2$ configuration, Cr$^{3+}$ has about 2.7 ${\rm \mu_B}$ local spin moment in both FM and AF states.
In the FM state, down-spin density is found at N-sites corresponding to the N --0.1 ${\rm \mu_B}$ local spin moment, and apparently, each pyrazine carries a considerable negative spin moment.
As in the bulk, the ligand contribution reduces the total magnetization to 2.00 ${\rm \mu_B}$/f.u., corresponding to the total $S = 1$ state.
In contrast, spin density almost vanishes at the N sites in the AF state, due to the counteracted spin-polarization induced by the AF Cr sublattice.

\begin{table}[t]
	\caption{Relative total energies ${\Delta \emph E}$ (meV/f.u.) for monolayer structures $\alpha$, $\beta$ and $\gamma$ in the FM and AF state. $\delta$-type structure converges to $\alpha$-type after relaxation.}{\label{table_ML}}
	\begin{tabular*}{0.48\textwidth}{@{\extracolsep{\fill}}ccccc}
		\hline
		states & {$\alpha$} & {$\beta$} & {$\gamma$} & {$\delta$} \\
		\hline
		FM & 0 & 182 & 302 & {$\rightarrow \alpha$} \\
		AF & 136 & 244 & 363 & \\
		\hline
	\end{tabular*}
\end{table}

\begin{figure}[h]
	\includegraphics[width=8.6cm]{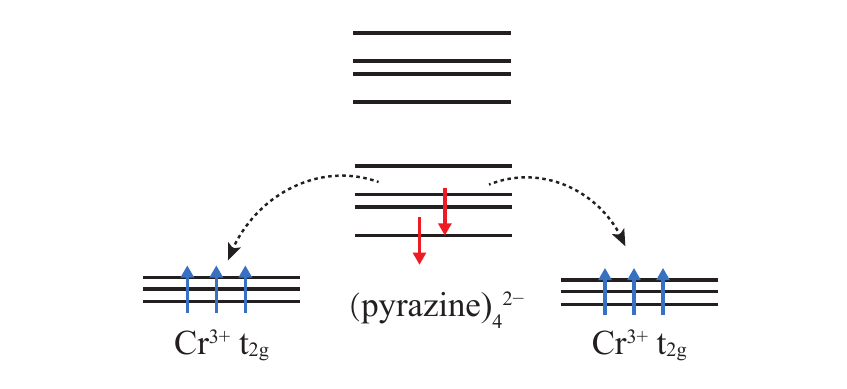}
	\caption{Schematic plot of the effective Cr-Cr FM interactions via the molecular orbitals of pyrazine ions in the monolayer CrCl$_2$(pyrazine)$_2$.}{\label{MO_illustrate}}
\end{figure}

%Molecular orbital
In order to study which orbitals the pyrazine bands near the Fermi level originate from, we carry out a calculation for an isolated pyrazine molecular.
The four energy levels near the Fermi level and the corresponding partial charge density are listed in Fig.\ \ref{monolayer}(e).
The energy levels fully occupied under the Fermi level should become deep valence bands in CrCl$_2$(pyrazine)$_2$ due to the higher chemical potential.
Among the empty energy levels above the Fermi level, the lowest one of 3.1 eV is of our concern.
This energy level, consisting of N 2$p_z$ and C 2$p_z$, is a molecular orbital which exists in organic cyclic compounds such as benzene or pyrazine.\cite{1975Electronic}
As discussed above, in CrCl$_2$(pyrazine)$_2$ one electron per Cr is transferred to pyrazines, and considering a lift of the Fermi level, the electron should come to this lowest unoccupied orbital.
Due to the two Cr and four pyrazines in a cell, there exist four such orbitals, which indicates a same origination for all the bands of concern.
To verify this, we plot the partial charge density for the bands near Fermi level in monolayer CrCl$_2$(pyrazine)$_2$ (see inserts of Fig.\ \ref{monolayer}(a) and (b)), and find all the charge density is similar to the isolated pyrazine molecular orbital except a small contribution from Cr.
A $p$-$d$ hybridization between pyrazine molecular orbitals and Cr $t_{2g}$ opens small energy splittings, as shown in Fig.\ \ref{monolayer}.
Then the large Cr polarization in the FM state gives rise to an exchange splitting, which results in a half-metallic character.
According to the Goodenough-Kanamori-Anderson (GKA) rules, a collinear Cr$^{3+}$-ligand-Cr$^{3+}$ superexchange should be AF.
Interestingly, what we find in CrCl$_2$(pyrazine)$_2$ is FM instead.
This is because, different from common cases in which no magnetization exists on the ligands, e.g. O$^{2-}$, here the pyrazine ligands have magnetic moments which give rise to the Cr-pyrazine coupling.
Owing to the half occupation of Cr $t_{2g}$ shell, the Cr-pyrazine direct exchange must be AF.
Therefore, the Cr-Cr FM coupling is stabilized, since FM allows electron itinerancy and gains much kinetic energy\cite{khomskii2014transition}, as shown in Fig.\ \ref{MO_illustrate}.

%\subsection{Doping}
\subsection{Carrier Doping}
The magnetism of 2D materials may be tuned by carrier or electrostatic doping\cite{jiang2018controlling,Cao2015doping}.
Hereby, as CrCl$_2$(pyrazine)$_2$ monolayer is a potential spintronic material, the possibility to enhance its FM coupling is of concern. For two Cr and four pyrazines in a cell, there exist four molecular orbitals near the Fermi level which are occupied by two electrons, and the polarization of the pyrazines by the Cr$^{3+}$ spin=3/2 gives rise to an exchange splitting with the down-spin levels being lower in energy, see Fig. 7. Thus, doping of $\pm$1 electron/f.u. ($\pm$2 electrons/cell) would completely occupy or deplete these four down-spin bands. Therefore, the electron doping from 0 to 1e/f.u. will increase the number of electrons hopping between pyrazine and Cr, and accordingly, gain more kinetic energy to further stabilize the FM state. But this would decrease the total magnetic moment as Cr and pyrazines have opposite spins. An even higher electron doping would occupy the four up-spin bands. Then the magnetic coupling between Cr and pyrazines will decrease, and this would reduce the Cr-Cr FM coupling. In contrast, the hole doping from 0 to --1e/f.u. will reduce the magnetic coupling between Cr and pyrazines, and in particular, the hole doping of --1e/f.u. will make the pyrazines formally nonmagnetic and then the tiny superexchange will give a weak Cr-Cr AF coupling. An even higher hole doping might deplete the deep valence bands which seems unrealistic and therefore is not discussed here.

\begin{table}[h]
	\caption{Relative total energies ${\Delta \emph E}$ (meV/f.u.), total and local spin moments (${\rm \mu_B}$) for the carrier doped FM and AF states.}{\label{table_doping}}
	\begin{tabular*}{0.48\textwidth}{@{\extracolsep{\fill}}ccrrrr}
		\hline
		&states&${\Delta \emph E}$&total&Cr&N\\
		\hline
		+1.5e&FM&0&1.41&2.69&--0.19 \\
		&AF&230&0.00&$\pm$2.62&$\mp$0.01 \\
		\hline
		+1.25e&FM&0&1.21&2.68&--0.19 \\
		&AF&233&0.00&$\pm$2.62&$\mp$0.01 \\
		\hline
		+1e&FM&0&1.00&2.67&--0.21 \\
		&AF&319&0.00&$\pm$2.64&$\mp$0.01 \\
		\hline
		+0.75e&FM&0&1.25&2.67&--0.18 \\
		&AF&296&0.00&$\pm$2.65&$\mp$0.01 \\
		\hline
		+0.5e&FM&0&1.50&2.68&--0.15 \\
		&AF&250&0.00&$\pm$2.66&$\mp$0.01 \\
		\hline
		+0.25e&FM&0&1.75&2.69&--0.12 \\
		&AF&193&0.00&$\pm$2.67&$\mp$0.01 \\
		\hline
		pure&FM&0&2.00&2.70&--0.10 \\
		&AF&136&0.00&$\pm$2.67&$\mp$0.01 \\
		\hline
		--0.25e&FM&0&2.25&2.71&--0.08 \\
		&AF&124&0.00&$\pm$2.71&$\mp$0.02 \\
		\hline
		--0.5e&FM&0&2.50&2.74&--0.06 \\
		&AF&98&0.00&$\pm$2.75&$\mp$0.02 \\
		\hline
		--0.75e&FM&0&2.75&2.76&--0.04 \\
		&AF&56&0.00&$\pm$2.77&$\mp$0.03 \\
		\hline
		--1e&FM&0&3.00&2.79&--0.02 \\
		&AF&--7&0.00&$\pm$2.79&$\mp$0.03 \\
		\hline
	\end{tabular*}
\end{table}

We now perform calculations for the carrier-doped CrCl$_2$(pyrazine)$_2$ monolayer, with the electron doping from 0 (pure) to 1.5 e/f.u., or hole doping from 0 to --1 e/f.u., both in a step of 0.25 e/f.u. The doping effect is simulated by adding or removing electrons in the unit cell, which is then neutralized by a background charge. As seen in Table 3, our calculations indeed confirm that the FM stability first increases but then decreases with the increasing electron doping from 0 to +1.5 e/f.u., and the FM ground state is most stable against the AF state by 319 meV/f.u. at the +1 e/f.u. doping. The local spin moment of Cr basically stays constant during the doping from 0 to +1.5 e/f.u., and the doped electrons fill up the ligands. In the FM ground state, the N atom (and the pyrazine) carries an increasing spin moment in the electron doping from 0 to 1 e/f.u., and the increasing down-spin density at the N atoms is clearly observed in Figs. 8(a) and 8(c).

A reverse process occurs for the hole doping, and the Cr-Cr FM coupling decreases for the hole doping from 0 to --1 e/f.u., as seen in Table 3. The doped holes into the ligands reduce the spin moment for the N atom (and the pyrazine), giving a lower down-spin density at the N atoms, see, e.g., Fig. 8(e). Note that for the hole doping of --0.75 e/f.u., there is 0.25 e/f.u. remaining in the pyrazine bands, and this still gives a strong Cr-Cr FM coupling (FM stability against AF by 56 meV/f.u.) as the magnetic coupling between Cr and pyrazines is quite effective. However, for the --1 e/f.u. doping, the pyrazine bands are completely depleted and become formally nonmagnetic, and then the above FM coupling is no longer effective, but there is now a weak superexchange Cr-Cr AF coupling (7 meV/f.u. AF stability against FM). Therefore, a FM-AF transition point could be very close to the --1 e/f.u. doping.

\begin{figure}[t]
\includegraphics[width=8cm]{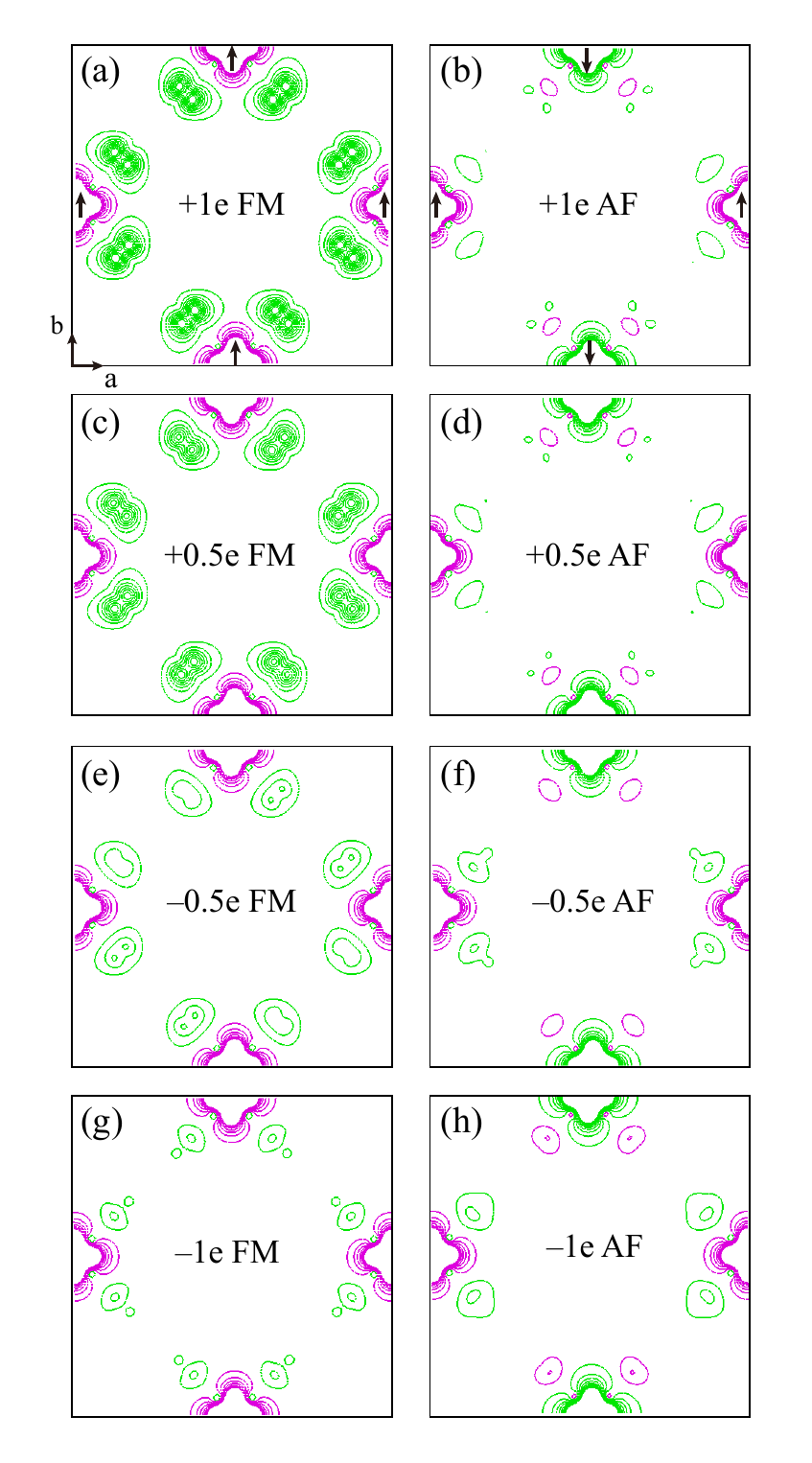}
\caption{Spin density plots of (a) 1 electron/f.u. FM, (b) 1 electron/f.u. AF, (c) 0.5 electron/f.u. FM, (d) 0.5 electron/f.u. AF, (e) 0.5 hole/f.u. FM, (f) 0.5 hole/f.u. AF, (h) 1 hole/f.u. FM and (h) 1 hole/f.u. AF states of monolayer CrCl$_2$(pyrazine)$_2$. The purple (green) lines are contours for up (down) spin density. The black arrows represent the spin moment directions of Cr.}{\label{doping}}
\end{figure}

%conclude
As seen above, we predict a significantly enhanced FM coupling in the electron-doped monolayer CrCl$_2$(pyrazine)$_2$ with the optimal doping of 1 e/f.u.
Moreover, a magnetic transition from FM to AF is predicted for hole doping very close to --1 e/f.u.
Then we establish the picture of the FM interaction via the spin-polarized molecular orbitals of the pyrazine ligands.
Thus, the CrCl$_2$(pyrazine)$_2$ monolayer, having a tunable FM half-metallicity, could be an appealing 2D spintronic material.

\section{Conclusions}
In summary, using density functional calculations and a magnetic picture, we confirm the half-metallicity in bulk CrCl$_2$(pyrazine)$_2$, and its strong intralayer FM and weak interlayer FM. These results agree well with the very recent experiments.
Our calculations show that the monolayer CrCl$_2$(pyrazine)$_2$ could be exfoliated from the bulk, and that its 2D FM half-metallicity remains robust.
Moreover, we provide a picture about the molecular orbitals of the pyrazine ligands and the magnetic couplings.
Based on this, we find that the electron doping can significantly enhance the FM coupling, but that the hole doping may even drive a FM-AF transition.
Therefore, the monolayer CrCl$_2$(pyrazine)$_2$ seems to be an appealing 2D spintronic material.
This work highlights the vital role of the organic ligands, and it suggests that 2D hybrid materials represent an interesting new platform with tunable electronic and magnetic properties which still need to be fully explored.

%\section{Acknowledgements}
This work was supported by the NSF of China (Grant No.11674064) and by the National Key Research and Development
Program of China (Grant No. 2016YFA0300700).

%\nocite{xyz}

\bibliographystyle{apsrev4-1}
\bibliography{Cr}% Produces the bibliography via BibTeX.

\begin{thebibliography}{46}%
\makeatletter
\providecommand \@ifxundefined [1]{%
 \@ifx{#1\undefined}
}%
\providecommand \@ifnum [1]{%
 \ifnum #1\expandafter \@firstoftwo
 \else \expandafter \@secondoftwo
 \fi
}%
\providecommand \@ifx [1]{%
 \ifx #1\expandafter \@firstoftwo
 \else \expandafter \@secondoftwo
 \fi
}%
\providecommand \natexlab [1]{#1}%
\providecommand \enquote  [1]{``#1''}%
\providecommand \bibnamefont  [1]{#1}%
\providecommand \bibfnamefont [1]{#1}%
\providecommand \citenamefont [1]{#1}%
\providecommand \href@noop [0]{\@secondoftwo}%
\providecommand \href [0]{\begingroup \@sanitize@url \@href}%
\providecommand \@href[1]{\@@startlink{#1}\@@href}%
\providecommand \@@href[1]{\endgroup#1\@@endlink}%
\providecommand \@sanitize@url [0]{\catcode `\\12\catcode `\$12\catcode
  `\&12\catcode `\#12\catcode `\^12\catcode `\_12\catcode `\%12\relax}%
\providecommand \@@startlink[1]{}%
\providecommand \@@endlink[0]{}%
\providecommand \url  [0]{\begingroup\@sanitize@url \@url }%
\providecommand \@url [1]{\endgroup\@href {#1}{\urlprefix }}%
\providecommand \urlprefix  [0]{URL }%
\providecommand \Eprint [0]{\href }%
\providecommand \doibase [0]{http://dx.doi.org/}%
\providecommand \selectlanguage [0]{\@gobble}%
\providecommand \bibinfo  [0]{\@secondoftwo}%
\providecommand \bibfield  [0]{\@secondoftwo}%
\providecommand \translation [1]{[#1]}%
\providecommand \BibitemOpen [0]{}%
\providecommand \bibitemStop [0]{}%
\providecommand \bibitemNoStop [0]{.\EOS\space}%
\providecommand \EOS [0]{\spacefactor3000\relax}%
\providecommand \BibitemShut  [1]{\csname bibitem#1\endcsname}%
\let\auto@bib@innerbib\@empty
%</preamble>
\bibitem [{\citenamefont {Wang}\ \emph {et~al.}(2018)\citenamefont {Wang},
  \citenamefont {Kong}, \citenamefont {Fan}, \citenamefont {Chen},
  \citenamefont {Zhu}, \citenamefont {Liu}, \citenamefont {Cao}, \citenamefont
  {Sun}, \citenamefont {Du}, \citenamefont {Schneeloch}, \citenamefont {Zhong},
  \citenamefont {Gu}, \citenamefont {Fu}, \citenamefont {Ding},\ and\
  \citenamefont {Gao}}]{wang2018evidence}%
  \BibitemOpen
  \bibfield  {author} {\bibinfo {author} {\bibfnamefont {D.}~\bibnamefont
  {Wang}}, \bibinfo {author} {\bibfnamefont {L.}~\bibnamefont {Kong}}, \bibinfo
  {author} {\bibfnamefont {P.}~\bibnamefont {Fan}}, \bibinfo {author}
  {\bibfnamefont {H.}~\bibnamefont {Chen}}, \bibinfo {author} {\bibfnamefont
  {S.}~\bibnamefont {Zhu}}, \bibinfo {author} {\bibfnamefont {W.}~\bibnamefont
  {Liu}}, \bibinfo {author} {\bibfnamefont {L.}~\bibnamefont {Cao}}, \bibinfo
  {author} {\bibfnamefont {Y.}~\bibnamefont {Sun}}, \bibinfo {author}
  {\bibfnamefont {S.}~\bibnamefont {Du}}, \bibinfo {author} {\bibfnamefont
  {J.}~\bibnamefont {Schneeloch}}, \bibinfo {author} {\bibfnamefont
  {R.}~\bibnamefont {Zhong}}, \bibinfo {author} {\bibfnamefont
  {G.}~\bibnamefont {Gu}}, \bibinfo {author} {\bibfnamefont {L.}~\bibnamefont
  {Fu}}, \bibinfo {author} {\bibfnamefont {H.}~\bibnamefont {Ding}}, \ and\
  \bibinfo {author} {\bibfnamefont {H.-J.}\ \bibnamefont {Gao}},\ }\href@noop
  {} {\bibfield  {journal} {\bibinfo  {journal} {Science}\ }\textbf {\bibinfo
  {volume} {362}},\ \bibinfo {pages} {333} (\bibinfo {year}
  {2018})}\BibitemShut {NoStop}%
\bibitem [{\citenamefont {Wu}\ \emph {et~al.}(2018)\citenamefont {Wu},
  \citenamefont {Fatemi}, \citenamefont {Gibson}, \citenamefont {Watanabe},
  \citenamefont {Taniguchi}, \citenamefont {Cava},\ and\ \citenamefont
  {Jarillo-Herrero}}]{wu2018observation}%
  \BibitemOpen
  \bibfield  {author} {\bibinfo {author} {\bibfnamefont {S.}~\bibnamefont
  {Wu}}, \bibinfo {author} {\bibfnamefont {V.}~\bibnamefont {Fatemi}}, \bibinfo
  {author} {\bibfnamefont {Q.~D.}\ \bibnamefont {Gibson}}, \bibinfo {author}
  {\bibfnamefont {K.}~\bibnamefont {Watanabe}}, \bibinfo {author}
  {\bibfnamefont {T.}~\bibnamefont {Taniguchi}}, \bibinfo {author}
  {\bibfnamefont {R.~J.}\ \bibnamefont {Cava}}, \ and\ \bibinfo {author}
  {\bibfnamefont {P.}~\bibnamefont {Jarillo-Herrero}},\ }\href@noop {}
  {\bibfield  {journal} {\bibinfo  {journal} {Science}\ }\textbf {\bibinfo
  {volume} {359}},\ \bibinfo {pages} {76} (\bibinfo {year} {2018})}\BibitemShut
  {NoStop}%
\bibitem [{\citenamefont {Tokura}\ \emph {et~al.}(2019)\citenamefont {Tokura},
  \citenamefont {Yasuda},\ and\ \citenamefont
  {Tsukazaki}}]{tokura2019magnetic}%
  \BibitemOpen
  \bibfield  {author} {\bibinfo {author} {\bibfnamefont {Y.}~\bibnamefont
  {Tokura}}, \bibinfo {author} {\bibfnamefont {K.}~\bibnamefont {Yasuda}}, \
  and\ \bibinfo {author} {\bibfnamefont {A.}~\bibnamefont {Tsukazaki}},\
  }\href@noop {} {\bibfield  {journal} {\bibinfo  {journal} {Nat. Rev. Phys.}\
  }\textbf {\bibinfo {volume} {1}},\ \bibinfo {pages} {126} (\bibinfo {year}
  {2019})}\BibitemShut {NoStop}%
\bibitem [{\citenamefont {Spaldin}\ and\ \citenamefont
  {Ramesh}(2019)}]{spaldin2019advances}%
  \BibitemOpen
  \bibfield  {author} {\bibinfo {author} {\bibfnamefont {N.~A.}\ \bibnamefont
  {Spaldin}}\ and\ \bibinfo {author} {\bibfnamefont {R.}~\bibnamefont
  {Ramesh}},\ }\href@noop {} {\bibfield  {journal} {\bibinfo  {journal} {Nat.
  Mater.}\ }\textbf {\bibinfo {volume} {18}},\ \bibinfo {pages} {203} (\bibinfo
  {year} {2019})}\BibitemShut {NoStop}%
\bibitem [{\citenamefont {Khomskii}(2014)}]{khomskii2014transition}%
  \BibitemOpen
  \bibfield  {author} {\bibinfo {author} {\bibfnamefont {D.}~\bibnamefont
  {Khomskii}},\ }\href@noop {} {}\ (\bibinfo  {publisher} {Cambridge University
  Press},\ \bibinfo {address} {Cambridge},\ \bibinfo {year} {2014})\BibitemShut
  {NoStop}%
\bibitem [{\citenamefont {Smejkal}\ \emph {et~al.}(2018)\citenamefont
  {Smejkal}, \citenamefont {Mokrousov}, \citenamefont {Yan},\ and\
  \citenamefont {MacDonald}}]{vsmejkal2018topological}%
  \BibitemOpen
  \bibfield  {author} {\bibinfo {author} {\bibfnamefont {L.}~\bibnamefont
  {Smejkal}}, \bibinfo {author} {\bibfnamefont {Y.}~\bibnamefont {Mokrousov}},
  \bibinfo {author} {\bibfnamefont {B.}~\bibnamefont {Yan}}, \ and\ \bibinfo
  {author} {\bibfnamefont {A.~H.}\ \bibnamefont {MacDonald}},\ }\href@noop {}
  {\bibfield  {journal} {\bibinfo  {journal} {Nat. Phys.}\ }\textbf {\bibinfo
  {volume} {14}},\ \bibinfo {pages} {242} (\bibinfo {year} {2018})}\BibitemShut
  {NoStop}%
\bibitem [{\citenamefont {Sun}\ \emph {et~al.}(2020)\citenamefont {Sun},
  \citenamefont {Yan}, \citenamefont {Liu}, \citenamefont {Xu}, \citenamefont
  {Cheng},\ and\ \citenamefont {Chen}}]{sun2020self}%
  \BibitemOpen
  \bibfield  {author} {\bibinfo {author} {\bibfnamefont {H.}~\bibnamefont
  {Sun}}, \bibinfo {author} {\bibfnamefont {Z.}~\bibnamefont {Yan}}, \bibinfo
  {author} {\bibfnamefont {F.}~\bibnamefont {Liu}}, \bibinfo {author}
  {\bibfnamefont {W.}~\bibnamefont {Xu}}, \bibinfo {author} {\bibfnamefont
  {F.}~\bibnamefont {Cheng}}, \ and\ \bibinfo {author} {\bibfnamefont
  {J.}~\bibnamefont {Chen}},\ }\href@noop {} {\bibfield  {journal} {\bibinfo
  {journal} {Adv. Mater.}\ }\textbf {\bibinfo {volume} {32}},\ \bibinfo {pages}
  {1806326} (\bibinfo {year} {2020})}\BibitemShut {NoStop}%
\bibitem [{\citenamefont {Manchon}\ \emph {et~al.}(2019)\citenamefont
  {Manchon}, \citenamefont {{\v{Z}}elezn{\`y}}, \citenamefont {Miron},
  \citenamefont {Jungwirth}, \citenamefont {Sinova}, \citenamefont {Thiaville},
  \citenamefont {Garello},\ and\ \citenamefont
  {Gambardella}}]{manchon2019current}%
  \BibitemOpen
  \bibfield  {author} {\bibinfo {author} {\bibfnamefont {A.}~\bibnamefont
  {Manchon}}, \bibinfo {author} {\bibfnamefont {J.}~\bibnamefont
  {{\v{Z}}elezn{\`y}}}, \bibinfo {author} {\bibfnamefont {I.~M.}\ \bibnamefont
  {Miron}}, \bibinfo {author} {\bibfnamefont {T.}~\bibnamefont {Jungwirth}},
  \bibinfo {author} {\bibfnamefont {J.}~\bibnamefont {Sinova}}, \bibinfo
  {author} {\bibfnamefont {A.}~\bibnamefont {Thiaville}}, \bibinfo {author}
  {\bibfnamefont {K.}~\bibnamefont {Garello}}, \ and\ \bibinfo {author}
  {\bibfnamefont {P.}~\bibnamefont {Gambardella}},\ }\href@noop {} {\bibfield
  {journal} {\bibinfo  {journal} {Rev. Mod. Phys.}\ }\textbf {\bibinfo {volume}
  {91}},\ \bibinfo {pages} {035004} (\bibinfo {year} {2019})}\BibitemShut
  {NoStop}%
\bibitem [{\citenamefont {Li}\ \emph {et~al.}(2016)\citenamefont {Li},
  \citenamefont {Wen}, \citenamefont {Cui}, \citenamefont {Zhou}, \citenamefont
  {Qian},\ and\ \citenamefont {Chen}}]{Li2016MOFreview}%
  \BibitemOpen
  \bibfield  {author} {\bibinfo {author} {\bibfnamefont {B.}~\bibnamefont
  {Li}}, \bibinfo {author} {\bibfnamefont {H.-M.}\ \bibnamefont {Wen}},
  \bibinfo {author} {\bibfnamefont {Y.}~\bibnamefont {Cui}}, \bibinfo {author}
  {\bibfnamefont {W.}~\bibnamefont {Zhou}}, \bibinfo {author} {\bibfnamefont
  {G.}~\bibnamefont {Qian}}, \ and\ \bibinfo {author} {\bibfnamefont
  {B.}~\bibnamefont {Chen}},\ }\href@noop {} {\bibfield  {journal} {\bibinfo
  {journal} {Adv. Mater.}\ }\textbf {\bibinfo {volume} {28}},\ \bibinfo {pages}
  {8819} (\bibinfo {year} {2016})}\BibitemShut {NoStop}%
\bibitem [{\citenamefont {Yuan}\ \emph {et~al.}(2018)\citenamefont {Yuan},
  \citenamefont {Feng}, \citenamefont {Wang}, \citenamefont {Pang},
  \citenamefont {Bosch}, \citenamefont {Lollar}, \citenamefont {Sun},
  \citenamefont {Qin}, \citenamefont {Yang}, \citenamefont {Zhang},
  \citenamefont {Wang}, \citenamefont {Zou}, \citenamefont {Zhang},
  \citenamefont {Zhang}, \citenamefont {Fang}, \citenamefont {Li},\ and\
  \citenamefont {Zhou}}]{yuan2018stable}%
  \BibitemOpen
  \bibfield  {author} {\bibinfo {author} {\bibfnamefont {S.}~\bibnamefont
  {Yuan}}, \bibinfo {author} {\bibfnamefont {L.}~\bibnamefont {Feng}}, \bibinfo
  {author} {\bibfnamefont {K.}~\bibnamefont {Wang}}, \bibinfo {author}
  {\bibfnamefont {J.}~\bibnamefont {Pang}}, \bibinfo {author} {\bibfnamefont
  {M.}~\bibnamefont {Bosch}}, \bibinfo {author} {\bibfnamefont
  {C.}~\bibnamefont {Lollar}}, \bibinfo {author} {\bibfnamefont
  {Y.}~\bibnamefont {Sun}}, \bibinfo {author} {\bibfnamefont {J.}~\bibnamefont
  {Qin}}, \bibinfo {author} {\bibfnamefont {X.}~\bibnamefont {Yang}}, \bibinfo
  {author} {\bibfnamefont {P.}~\bibnamefont {Zhang}}, \bibinfo {author}
  {\bibfnamefont {Q.}~\bibnamefont {Wang}}, \bibinfo {author} {\bibfnamefont
  {L.}~\bibnamefont {Zou}}, \bibinfo {author} {\bibfnamefont {Y.}~\bibnamefont
  {Zhang}}, \bibinfo {author} {\bibfnamefont {L.}~\bibnamefont {Zhang}},
  \bibinfo {author} {\bibfnamefont {Y.}~\bibnamefont {Fang}}, \bibinfo {author}
  {\bibfnamefont {J.}~\bibnamefont {Li}}, \ and\ \bibinfo {author}
  {\bibfnamefont {H.-C.}\ \bibnamefont {Zhou}},\ }\href@noop {} {\bibfield
  {journal} {\bibinfo  {journal} {Adv. Mater.}\ }\textbf {\bibinfo {volume}
  {30}},\ \bibinfo {pages} {1704303} (\bibinfo {year} {2018})}\BibitemShut
  {NoStop}%
\bibitem [{\citenamefont {Coronado}(2020)}]{coronado2020molecular}%
  \BibitemOpen
  \bibfield  {author} {\bibinfo {author} {\bibfnamefont {E.}~\bibnamefont
  {Coronado}},\ }\href@noop {} {\bibfield  {journal} {\bibinfo  {journal} {Nat.
  Rev. Mater.}\ }\textbf {\bibinfo {volume} {5}},\ \bibinfo {pages} {87}
  (\bibinfo {year} {2020})}\BibitemShut {NoStop}%
\bibitem [{\citenamefont {Stroppa}\ \emph {et~al.}(2014)\citenamefont
  {Stroppa}, \citenamefont {Di~Sante}, \citenamefont {Barone}, \citenamefont
  {Bokdam}, \citenamefont {Kresse}, \citenamefont {Franchini}, \citenamefont
  {Whangbo},\ and\ \citenamefont {Picozzi}}]{Stroppa2014ferroelectric}%
  \BibitemOpen
  \bibfield  {author} {\bibinfo {author} {\bibfnamefont {A.}~\bibnamefont
  {Stroppa}}, \bibinfo {author} {\bibfnamefont {D.}~\bibnamefont {Di~Sante}},
  \bibinfo {author} {\bibfnamefont {P.}~\bibnamefont {Barone}}, \bibinfo
  {author} {\bibfnamefont {M.}~\bibnamefont {Bokdam}}, \bibinfo {author}
  {\bibfnamefont {G.}~\bibnamefont {Kresse}}, \bibinfo {author} {\bibfnamefont
  {C.}~\bibnamefont {Franchini}}, \bibinfo {author} {\bibfnamefont {M.~H.}\
  \bibnamefont {Whangbo}}, \ and\ \bibinfo {author} {\bibfnamefont
  {S.}~\bibnamefont {Picozzi}},\ }\href@noop {} {\bibfield  {journal} {\bibinfo
   {journal} {Nat. Commun.}\ }\textbf {\bibinfo {volume} {5}},\ \bibinfo
  {pages} {5900} (\bibinfo {year} {2014})}\BibitemShut {NoStop}%
\bibitem [{\citenamefont {Darago}\ \emph {et~al.}(2015)\citenamefont {Darago},
  \citenamefont {Aubrey}, \citenamefont {Yu}, \citenamefont {Gonzalez},\ and\
  \citenamefont {Long}}]{darago2015electronic}%
  \BibitemOpen
  \bibfield  {author} {\bibinfo {author} {\bibfnamefont {L.~E.}\ \bibnamefont
  {Darago}}, \bibinfo {author} {\bibfnamefont {M.~L.}\ \bibnamefont {Aubrey}},
  \bibinfo {author} {\bibfnamefont {C.~J.}\ \bibnamefont {Yu}}, \bibinfo
  {author} {\bibfnamefont {M.~I.}\ \bibnamefont {Gonzalez}}, \ and\ \bibinfo
  {author} {\bibfnamefont {J.~R.}\ \bibnamefont {Long}},\ }\href@noop {}
  {\bibfield  {journal} {\bibinfo  {journal} {J. Am. Chem. Soc.}\ }\textbf
  {\bibinfo {volume} {137}},\ \bibinfo {pages} {15703} (\bibinfo {year}
  {2015})}\BibitemShut {NoStop}%
\bibitem [{\citenamefont {Ma}\ \emph {et~al.}(2019)\citenamefont {Ma},
  \citenamefont {Suturina}, \citenamefont {Rouzi{\`e}res}, \citenamefont
  {Platunov}, \citenamefont {Wilhelm}, \citenamefont {Rogalev}, \citenamefont
  {Cl{\'e}rac},\ and\ \citenamefont {Dechambenoit}}]{ma2019using}%
  \BibitemOpen
  \bibfield  {author} {\bibinfo {author} {\bibfnamefont {X.}~\bibnamefont
  {Ma}}, \bibinfo {author} {\bibfnamefont {E.~A.}\ \bibnamefont {Suturina}},
  \bibinfo {author} {\bibfnamefont {M.}~\bibnamefont {Rouzi{\`e}res}}, \bibinfo
  {author} {\bibfnamefont {M.}~\bibnamefont {Platunov}}, \bibinfo {author}
  {\bibfnamefont {F.}~\bibnamefont {Wilhelm}}, \bibinfo {author} {\bibfnamefont
  {A.}~\bibnamefont {Rogalev}}, \bibinfo {author} {\bibfnamefont
  {R.}~\bibnamefont {Cl{\'e}rac}}, \ and\ \bibinfo {author} {\bibfnamefont
  {P.}~\bibnamefont {Dechambenoit}},\ }\href@noop {} {\bibfield  {journal}
  {\bibinfo  {journal} {{J}. {A}m. {C}hem. {S}oc.}\ }\textbf {\bibinfo {volume}
  {141}},\ \bibinfo {pages} {7721} (\bibinfo {year} {2019})}\BibitemShut
  {NoStop}%
\bibitem [{\citenamefont {Li}\ \emph {et~al.}(2017)\citenamefont {Li},
  \citenamefont {Kim}, \citenamefont {Jin}, \citenamefont {Ye}, \citenamefont
  {Qiu}, \citenamefont {Felipe}, \citenamefont {Shi}, \citenamefont {Chen},
  \citenamefont {Zhang}, \citenamefont {Yang}, \citenamefont {Watanabe},
  \citenamefont {Taniguchi}, \citenamefont {Ren}, \citenamefont {Louie},
  \citenamefont {Chen}, \citenamefont {Zhang},\ and\ \citenamefont
  {Wang}}]{li2017direct}%
  \BibitemOpen
  \bibfield  {author} {\bibinfo {author} {\bibfnamefont {L.}~\bibnamefont
  {Li}}, \bibinfo {author} {\bibfnamefont {J.}~\bibnamefont {Kim}}, \bibinfo
  {author} {\bibfnamefont {C.}~\bibnamefont {Jin}}, \bibinfo {author}
  {\bibfnamefont {G.~J.}\ \bibnamefont {Ye}}, \bibinfo {author} {\bibfnamefont
  {D.~Y.}\ \bibnamefont {Qiu}}, \bibinfo {author} {\bibfnamefont
  {H.}~\bibnamefont {Felipe}}, \bibinfo {author} {\bibfnamefont
  {Z.}~\bibnamefont {Shi}}, \bibinfo {author} {\bibfnamefont {L.}~\bibnamefont
  {Chen}}, \bibinfo {author} {\bibfnamefont {Z.}~\bibnamefont {Zhang}},
  \bibinfo {author} {\bibfnamefont {F.}~\bibnamefont {Yang}}, \bibinfo {author}
  {\bibfnamefont {K.}~\bibnamefont {Watanabe}}, \bibinfo {author}
  {\bibfnamefont {T.}~\bibnamefont {Taniguchi}}, \bibinfo {author}
  {\bibfnamefont {W.}~\bibnamefont {Ren}}, \bibinfo {author} {\bibfnamefont
  {S.~G.}\ \bibnamefont {Louie}}, \bibinfo {author} {\bibfnamefont {X.~H.}\
  \bibnamefont {Chen}}, \bibinfo {author} {\bibfnamefont {Y.}~\bibnamefont
  {Zhang}}, \ and\ \bibinfo {author} {\bibfnamefont {F.}~\bibnamefont {Wang}},\
  }\href@noop {} {\bibfield  {journal} {\bibinfo  {journal} {{N}at.
  {N}anotechnol.}\ }\textbf {\bibinfo {volume} {12}},\ \bibinfo {pages} {21}
  (\bibinfo {year} {2017})}\BibitemShut {NoStop}%
\bibitem [{\citenamefont {Huang}\ \emph {et~al.}(2017)\citenamefont {Huang},
  \citenamefont {Clark}, \citenamefont {Navarro-Moratalla}, \citenamefont
  {Klein}, \citenamefont {Cheng}, \citenamefont {Seyler}, \citenamefont
  {Zhong}, \citenamefont {Schmidgall}, \citenamefont {McGuire}, \citenamefont
  {Cobden}, \citenamefont {Yao}, \citenamefont {Xiao}, \citenamefont
  {Jarillo-Herrero},\ and\ \citenamefont {Xu}}]{huang2017layer}%
  \BibitemOpen
  \bibfield  {author} {\bibinfo {author} {\bibfnamefont {B.}~\bibnamefont
  {Huang}}, \bibinfo {author} {\bibfnamefont {G.}~\bibnamefont {Clark}},
  \bibinfo {author} {\bibfnamefont {E.}~\bibnamefont {Navarro-Moratalla}},
  \bibinfo {author} {\bibfnamefont {D.~R.}\ \bibnamefont {Klein}}, \bibinfo
  {author} {\bibfnamefont {R.}~\bibnamefont {Cheng}}, \bibinfo {author}
  {\bibfnamefont {K.~L.}\ \bibnamefont {Seyler}}, \bibinfo {author}
  {\bibfnamefont {D.}~\bibnamefont {Zhong}}, \bibinfo {author} {\bibfnamefont
  {E.}~\bibnamefont {Schmidgall}}, \bibinfo {author} {\bibfnamefont {M.~A.}\
  \bibnamefont {McGuire}}, \bibinfo {author} {\bibfnamefont {D.~H.}\
  \bibnamefont {Cobden}}, \bibinfo {author} {\bibfnamefont {W.}~\bibnamefont
  {Yao}}, \bibinfo {author} {\bibfnamefont {D.}~\bibnamefont {Xiao}}, \bibinfo
  {author} {\bibfnamefont {P.}~\bibnamefont {Jarillo-Herrero}}, \ and\ \bibinfo
  {author} {\bibfnamefont {X.}~\bibnamefont {Xu}},\ }\href@noop {} {\bibfield
  {journal} {\bibinfo  {journal} {{N}ature}\ }\textbf {\bibinfo {volume}
  {546}},\ \bibinfo {pages} {270} (\bibinfo {year} {2017})}\BibitemShut
  {NoStop}%
\bibitem [{\citenamefont {Gong}\ \emph {et~al.}(2017)\citenamefont {Gong},
  \citenamefont {Li}, \citenamefont {Li}, \citenamefont {Ji}, \citenamefont
  {Stern}, \citenamefont {Xia}, \citenamefont {Cao}, \citenamefont {Bao},
  \citenamefont {Wang}, \citenamefont {Wang}, \citenamefont {Qiu},
  \citenamefont {Cava}, \citenamefont {Louie}, \citenamefont {Xia},\ and\
  \citenamefont {Zhang}}]{gong2017discovery}%
  \BibitemOpen
  \bibfield  {author} {\bibinfo {author} {\bibfnamefont {C.}~\bibnamefont
  {Gong}}, \bibinfo {author} {\bibfnamefont {L.}~\bibnamefont {Li}}, \bibinfo
  {author} {\bibfnamefont {Z.}~\bibnamefont {Li}}, \bibinfo {author}
  {\bibfnamefont {H.}~\bibnamefont {Ji}}, \bibinfo {author} {\bibfnamefont
  {A.}~\bibnamefont {Stern}}, \bibinfo {author} {\bibfnamefont
  {Y.}~\bibnamefont {Xia}}, \bibinfo {author} {\bibfnamefont {T.}~\bibnamefont
  {Cao}}, \bibinfo {author} {\bibfnamefont {W.}~\bibnamefont {Bao}}, \bibinfo
  {author} {\bibfnamefont {C.}~\bibnamefont {Wang}}, \bibinfo {author}
  {\bibfnamefont {Y.}~\bibnamefont {Wang}}, \bibinfo {author} {\bibfnamefont
  {Z.~Q.}\ \bibnamefont {Qiu}}, \bibinfo {author} {\bibfnamefont {R.~J.}\
  \bibnamefont {Cava}}, \bibinfo {author} {\bibfnamefont {S.~G.}\ \bibnamefont
  {Louie}}, \bibinfo {author} {\bibfnamefont {J.}~\bibnamefont {Xia}}, \ and\
  \bibinfo {author} {\bibfnamefont {X.}~\bibnamefont {Zhang}},\ }\href@noop {}
  {\bibfield  {journal} {\bibinfo  {journal} {{N}ature}\ }\textbf {\bibinfo
  {volume} {546}},\ \bibinfo {pages} {265} (\bibinfo {year}
  {2017})}\BibitemShut {NoStop}%
\bibitem [{\citenamefont {Deng}\ \emph {et~al.}(2018)\citenamefont {Deng},
  \citenamefont {Yu}, \citenamefont {Song}, \citenamefont {Zhang},
  \citenamefont {Wang}, \citenamefont {Sun}, \citenamefont {Yi}, \citenamefont
  {Wu}, \citenamefont {Wu}, \citenamefont {Zhu}, \citenamefont {Wang},
  \citenamefont {Chen},\ and\ \citenamefont {Zhang}}]{deng2018gate}%
  \BibitemOpen
  \bibfield  {author} {\bibinfo {author} {\bibfnamefont {Y.}~\bibnamefont
  {Deng}}, \bibinfo {author} {\bibfnamefont {Y.}~\bibnamefont {Yu}}, \bibinfo
  {author} {\bibfnamefont {Y.}~\bibnamefont {Song}}, \bibinfo {author}
  {\bibfnamefont {J.}~\bibnamefont {Zhang}}, \bibinfo {author} {\bibfnamefont
  {N.~Z.}\ \bibnamefont {Wang}}, \bibinfo {author} {\bibfnamefont
  {Z.}~\bibnamefont {Sun}}, \bibinfo {author} {\bibfnamefont {Y.}~\bibnamefont
  {Yi}}, \bibinfo {author} {\bibfnamefont {Y.~Z.}\ \bibnamefont {Wu}}, \bibinfo
  {author} {\bibfnamefont {S.}~\bibnamefont {Wu}}, \bibinfo {author}
  {\bibfnamefont {J.}~\bibnamefont {Zhu}}, \bibinfo {author} {\bibfnamefont
  {J.}~\bibnamefont {Wang}}, \bibinfo {author} {\bibfnamefont {X.~H.}\
  \bibnamefont {Chen}}, \ and\ \bibinfo {author} {\bibfnamefont
  {Y.}~\bibnamefont {Zhang}},\ }\href@noop {} {\bibfield  {journal} {\bibinfo
  {journal} {{N}ature}\ }\textbf {\bibinfo {volume} {563}},\ \bibinfo {pages}
  {94} (\bibinfo {year} {2018})}\BibitemShut {NoStop}%
\bibitem [{\citenamefont {Fei}\ \emph {et~al.}(2018)\citenamefont {Fei},
  \citenamefont {Huang}, \citenamefont {Malinowski}, \citenamefont {Wang},
  \citenamefont {Song}, \citenamefont {Sanchez}, \citenamefont {Yao},
  \citenamefont {Xiao}, \citenamefont {Zhu}, \citenamefont {May}, \citenamefont
  {Wu}, \citenamefont {Cobden}, \citenamefont {Chu},\ and\ \citenamefont
  {Xu}}]{fei2018Fe3GeTe2}%
  \BibitemOpen
  \bibfield  {author} {\bibinfo {author} {\bibfnamefont {Z.}~\bibnamefont
  {Fei}}, \bibinfo {author} {\bibfnamefont {B.}~\bibnamefont {Huang}}, \bibinfo
  {author} {\bibfnamefont {P.}~\bibnamefont {Malinowski}}, \bibinfo {author}
  {\bibfnamefont {W.}~\bibnamefont {Wang}}, \bibinfo {author} {\bibfnamefont
  {T.}~\bibnamefont {Song}}, \bibinfo {author} {\bibfnamefont {J.}~\bibnamefont
  {Sanchez}}, \bibinfo {author} {\bibfnamefont {W.}~\bibnamefont {Yao}},
  \bibinfo {author} {\bibfnamefont {D.}~\bibnamefont {Xiao}}, \bibinfo {author}
  {\bibfnamefont {X.}~\bibnamefont {Zhu}}, \bibinfo {author} {\bibfnamefont
  {A.~F.}\ \bibnamefont {May}}, \bibinfo {author} {\bibfnamefont
  {W.}~\bibnamefont {Wu}}, \bibinfo {author} {\bibfnamefont {D.~H.}\
  \bibnamefont {Cobden}}, \bibinfo {author} {\bibfnamefont {J.-H.}\
  \bibnamefont {Chu}}, \ and\ \bibinfo {author} {\bibfnamefont
  {X.}~\bibnamefont {Xu}},\ }\href@noop {} {\bibfield  {journal} {\bibinfo
  {journal} {{N}at. {M}ater.}\ }\textbf {\bibinfo {volume} {17}},\ \bibinfo
  {pages} {778} (\bibinfo {year} {2018})}\BibitemShut {NoStop}%
\bibitem [{\citenamefont {Lee}\ \emph {et~al.}(2016)\citenamefont {Lee},
  \citenamefont {Lee}, \citenamefont {Ryoo}, \citenamefont {Kang},
  \citenamefont {Kim}, \citenamefont {Kim}, \citenamefont {Park}, \citenamefont
  {Park},\ and\ \citenamefont {Cheong}}]{Lee2016FePS3}%
  \BibitemOpen
  \bibfield  {author} {\bibinfo {author} {\bibfnamefont {J.~U.}\ \bibnamefont
  {Lee}}, \bibinfo {author} {\bibfnamefont {S.}~\bibnamefont {Lee}}, \bibinfo
  {author} {\bibfnamefont {J.~H.}\ \bibnamefont {Ryoo}}, \bibinfo {author}
  {\bibfnamefont {S.}~\bibnamefont {Kang}}, \bibinfo {author} {\bibfnamefont
  {T.~Y.}\ \bibnamefont {Kim}}, \bibinfo {author} {\bibfnamefont
  {P.}~\bibnamefont {Kim}}, \bibinfo {author} {\bibfnamefont {C.~H.}\
  \bibnamefont {Park}}, \bibinfo {author} {\bibfnamefont {J.~G.}\ \bibnamefont
  {Park}}, \ and\ \bibinfo {author} {\bibfnamefont {H.}~\bibnamefont
  {Cheong}},\ }\href@noop {} {\bibfield  {journal} {\bibinfo  {journal} {{N}ano
  {L}ett.}\ }\textbf {\bibinfo {volume} {16}},\ \bibinfo {pages} {7433}
  (\bibinfo {year} {2016})}\BibitemShut {NoStop}%
\bibitem [{\citenamefont {Lin}\ \emph {et~al.}(2016)\citenamefont {Lin},
  \citenamefont {Zhuang}, \citenamefont {Yan}, \citenamefont {Ward},
  \citenamefont {Puretzky}, \citenamefont {Rouleau}, \citenamefont {Gai},
  \citenamefont {Liang}, \citenamefont {Meunier}, \citenamefont {Sumpter},
  \citenamefont {Ganesh}, \citenamefont {Kent}, \citenamefont {Geohegan},
  \citenamefont {Mandrus},\ and\ \citenamefont {Xiao}}]{Lin2016CrSiTe3}%
  \BibitemOpen
  \bibfield  {author} {\bibinfo {author} {\bibfnamefont {M.~W.}\ \bibnamefont
  {Lin}}, \bibinfo {author} {\bibfnamefont {H.~L.}\ \bibnamefont {Zhuang}},
  \bibinfo {author} {\bibfnamefont {J.}~\bibnamefont {Yan}}, \bibinfo {author}
  {\bibfnamefont {T.~Z.}\ \bibnamefont {Ward}}, \bibinfo {author}
  {\bibfnamefont {A.~A.}\ \bibnamefont {Puretzky}}, \bibinfo {author}
  {\bibfnamefont {C.~M.}\ \bibnamefont {Rouleau}}, \bibinfo {author}
  {\bibfnamefont {Z.}~\bibnamefont {Gai}}, \bibinfo {author} {\bibfnamefont
  {L.}~\bibnamefont {Liang}}, \bibinfo {author} {\bibfnamefont
  {V.}~\bibnamefont {Meunier}}, \bibinfo {author} {\bibfnamefont {B.~G.}\
  \bibnamefont {Sumpter}}, \bibinfo {author} {\bibfnamefont {P.}~\bibnamefont
  {Ganesh}}, \bibinfo {author} {\bibfnamefont {P.~R.~C.}\ \bibnamefont {Kent}},
  \bibinfo {author} {\bibfnamefont {D.~B.}\ \bibnamefont {Geohegan}}, \bibinfo
  {author} {\bibfnamefont {D.~G.}\ \bibnamefont {Mandrus}}, \ and\ \bibinfo
  {author} {\bibfnamefont {K.}~\bibnamefont {Xiao}},\ }\href@noop {} {\bibfield
   {journal} {\bibinfo  {journal} {{J}. {M}ater. {C}hem. {C}}\ }\textbf
  {\bibinfo {volume} {4}},\ \bibinfo {pages} {315} (\bibinfo {year}
  {2016})}\BibitemShut {NoStop}%
\bibitem [{\citenamefont {Bonilla}\ \emph {et~al.}(2018)\citenamefont
  {Bonilla}, \citenamefont {Kolekar}, \citenamefont {Ma}, \citenamefont {Diaz},
  \citenamefont {Kalappattil}, \citenamefont {Das}, \citenamefont {Eggers},
  \citenamefont {Gutierrez}, \citenamefont {Phan},\ and\ \citenamefont
  {Batzill}}]{Bonilla2018VSe2}%
  \BibitemOpen
  \bibfield  {author} {\bibinfo {author} {\bibfnamefont {M.}~\bibnamefont
  {Bonilla}}, \bibinfo {author} {\bibfnamefont {S.}~\bibnamefont {Kolekar}},
  \bibinfo {author} {\bibfnamefont {Y.}~\bibnamefont {Ma}}, \bibinfo {author}
  {\bibfnamefont {H.~C.}\ \bibnamefont {Diaz}}, \bibinfo {author}
  {\bibfnamefont {V.}~\bibnamefont {Kalappattil}}, \bibinfo {author}
  {\bibfnamefont {R.}~\bibnamefont {Das}}, \bibinfo {author} {\bibfnamefont
  {T.}~\bibnamefont {Eggers}}, \bibinfo {author} {\bibfnamefont {H.~R.}\
  \bibnamefont {Gutierrez}}, \bibinfo {author} {\bibfnamefont {M.~H.}\
  \bibnamefont {Phan}}, \ and\ \bibinfo {author} {\bibfnamefont
  {M.}~\bibnamefont {Batzill}},\ }\href@noop {} {\bibfield  {journal} {\bibinfo
   {journal} {{N}at. {N}anotechnol.}\ }\textbf {\bibinfo {volume} {13}},\
  \bibinfo {pages} {289} (\bibinfo {year} {2018})}\BibitemShut {NoStop}%
\bibitem [{\citenamefont {Kazim}\ \emph {et~al.}(2020)\citenamefont {Kazim},
  \citenamefont {Ali}, \citenamefont {Palleschi}, \citenamefont {D'Olimpio},
  \citenamefont {Mastrippolito}, \citenamefont {Politano}, \citenamefont
  {Gunnella}, \citenamefont {Di~Cicco}, \citenamefont {Renzelli}, \citenamefont
  {Moccia}, \citenamefont {Cacioppo}, \citenamefont {Alfonsetti}, \citenamefont
  {Strychalska-Nowak}, \citenamefont {Klimczuk}, \citenamefont {Cava},\ and\
  \citenamefont {Ottaviano}}]{kazim2020mechanical}%
  \BibitemOpen
  \bibfield  {author} {\bibinfo {author} {\bibfnamefont {S.}~\bibnamefont
  {Kazim}}, \bibinfo {author} {\bibfnamefont {M.}~\bibnamefont {Ali}}, \bibinfo
  {author} {\bibfnamefont {S.}~\bibnamefont {Palleschi}}, \bibinfo {author}
  {\bibfnamefont {G.}~\bibnamefont {D'Olimpio}}, \bibinfo {author}
  {\bibfnamefont {D.}~\bibnamefont {Mastrippolito}}, \bibinfo {author}
  {\bibfnamefont {A.}~\bibnamefont {Politano}}, \bibinfo {author}
  {\bibfnamefont {R.}~\bibnamefont {Gunnella}}, \bibinfo {author}
  {\bibfnamefont {A.}~\bibnamefont {Di~Cicco}}, \bibinfo {author}
  {\bibfnamefont {M.}~\bibnamefont {Renzelli}}, \bibinfo {author}
  {\bibfnamefont {G.}~\bibnamefont {Moccia}}, \bibinfo {author} {\bibfnamefont
  {O.~A.}\ \bibnamefont {Cacioppo}}, \bibinfo {author} {\bibfnamefont
  {R.}~\bibnamefont {Alfonsetti}}, \bibinfo {author} {\bibfnamefont
  {J.}~\bibnamefont {Strychalska-Nowak}}, \bibinfo {author} {\bibfnamefont
  {T.}~\bibnamefont {Klimczuk}}, \bibinfo {author} {\bibfnamefont {R.~J.}\
  \bibnamefont {Cava}}, \ and\ \bibinfo {author} {\bibfnamefont
  {L.}~\bibnamefont {Ottaviano}},\ }\href@noop {} {\bibfield  {journal}
  {\bibinfo  {journal} {{N}anotechnology}\ }\textbf {\bibinfo {volume} {31}},\
  \bibinfo {pages} {395706} (\bibinfo {year} {2020})}\BibitemShut {NoStop}%
\bibitem [{\citenamefont {Serri}\ \emph {et~al.}(2020)\citenamefont {Serri},
  \citenamefont {Cucinotta}, \citenamefont {Poggini}, \citenamefont {Serrano},
  \citenamefont {Sainctavit}, \citenamefont {Strychalska-Nowak}, \citenamefont
  {Politano}, \citenamefont {Bonaccorso}, \citenamefont {Caneschi},
  \citenamefont {Cava}, \citenamefont {Sessoli}, \citenamefont {Ottaviano},
  \citenamefont {Klimczuk}, \citenamefont {Pellegrini},\ and\ \citenamefont
  {Mannini}}]{serri2020enhancement}%
  \BibitemOpen
  \bibfield  {author} {\bibinfo {author} {\bibfnamefont {M.}~\bibnamefont
  {Serri}}, \bibinfo {author} {\bibfnamefont {G.}~\bibnamefont {Cucinotta}},
  \bibinfo {author} {\bibfnamefont {L.}~\bibnamefont {Poggini}}, \bibinfo
  {author} {\bibfnamefont {G.}~\bibnamefont {Serrano}}, \bibinfo {author}
  {\bibfnamefont {P.}~\bibnamefont {Sainctavit}}, \bibinfo {author}
  {\bibfnamefont {J.}~\bibnamefont {Strychalska-Nowak}}, \bibinfo {author}
  {\bibfnamefont {A.}~\bibnamefont {Politano}}, \bibinfo {author}
  {\bibfnamefont {F.}~\bibnamefont {Bonaccorso}}, \bibinfo {author}
  {\bibfnamefont {A.}~\bibnamefont {Caneschi}}, \bibinfo {author}
  {\bibfnamefont {R.~J.}\ \bibnamefont {Cava}}, \bibinfo {author}
  {\bibfnamefont {R.}~\bibnamefont {Sessoli}}, \bibinfo {author} {\bibfnamefont
  {L.}~\bibnamefont {Ottaviano}}, \bibinfo {author} {\bibfnamefont
  {T.}~\bibnamefont {Klimczuk}}, \bibinfo {author} {\bibfnamefont
  {V.}~\bibnamefont {Pellegrini}}, \ and\ \bibinfo {author} {\bibfnamefont
  {M.}~\bibnamefont {Mannini}},\ }\href@noop {} {\bibfield  {journal} {\bibinfo
   {journal} {{A}dv. {M}ater.}\ }\textbf {\bibinfo {volume} {32}},\ \bibinfo
  {pages} {2000566} (\bibinfo {year} {2020})}\BibitemShut {NoStop}%
\bibitem [{\citenamefont {Lado}\ and\ \citenamefont
  {Fern\'andez-Rossier}(2017)}]{Lado2017anisotropy}%
  \BibitemOpen
  \bibfield  {author} {\bibinfo {author} {\bibfnamefont {J.~L.}\ \bibnamefont
  {Lado}}\ and\ \bibinfo {author} {\bibfnamefont {J.}~\bibnamefont
  {Fern\'andez-Rossier}},\ }\href@noop {} {\bibfield  {journal} {\bibinfo
  {journal} {{2D} {M}ater.}\ }\textbf {\bibinfo {volume} {4}},\ \bibinfo
  {pages} {035002} (\bibinfo {year} {2017})}\BibitemShut {NoStop}%
\bibitem [{\citenamefont {Kim}\ \emph {et~al.}(2019)\citenamefont {Kim},
  \citenamefont {Kim}, \citenamefont {Ko}, \citenamefont {Seo}, \citenamefont
  {Kim}, \citenamefont {Jang}, \citenamefont {Kim}, \citenamefont {Kim},
  \citenamefont {Cheong},\ and\ \citenamefont {Park}}]{Kim2019LS}%
  \BibitemOpen
  \bibfield  {author} {\bibinfo {author} {\bibfnamefont {D.-H.}\ \bibnamefont
  {Kim}}, \bibinfo {author} {\bibfnamefont {K.}~\bibnamefont {Kim}}, \bibinfo
  {author} {\bibfnamefont {K.-T.}\ \bibnamefont {Ko}}, \bibinfo {author}
  {\bibfnamefont {J.}~\bibnamefont {Seo}}, \bibinfo {author} {\bibfnamefont
  {J.~S.}\ \bibnamefont {Kim}}, \bibinfo {author} {\bibfnamefont {T.-H.}\
  \bibnamefont {Jang}}, \bibinfo {author} {\bibfnamefont {Y.}~\bibnamefont
  {Kim}}, \bibinfo {author} {\bibfnamefont {J.-Y.}\ \bibnamefont {Kim}},
  \bibinfo {author} {\bibfnamefont {S.-W.}\ \bibnamefont {Cheong}}, \ and\
  \bibinfo {author} {\bibfnamefont {J.-H.}\ \bibnamefont {Park}},\ }\href@noop
  {} {\bibfield  {journal} {\bibinfo  {journal} {{P}hys. {R}ev. {L}ett.}\
  }\textbf {\bibinfo {volume} {122}},\ \bibinfo {pages} {207201} (\bibinfo
  {year} {2019})}\BibitemShut {NoStop}%
\bibitem [{\citenamefont {Jiang}\ \emph {et~al.}(2019)\citenamefont {Jiang},
  \citenamefont {Wang}, \citenamefont {Chen}, \citenamefont {Zhong},
  \citenamefont {Yuan}, \citenamefont {Lu},\ and\ \citenamefont
  {Ji}}]{Jiang2019stacking}%
  \BibitemOpen
  \bibfield  {author} {\bibinfo {author} {\bibfnamefont {P.}~\bibnamefont
  {Jiang}}, \bibinfo {author} {\bibfnamefont {C.}~\bibnamefont {Wang}},
  \bibinfo {author} {\bibfnamefont {D.}~\bibnamefont {Chen}}, \bibinfo {author}
  {\bibfnamefont {Z.}~\bibnamefont {Zhong}}, \bibinfo {author} {\bibfnamefont
  {Z.}~\bibnamefont {Yuan}}, \bibinfo {author} {\bibfnamefont {Z.-Y.}\
  \bibnamefont {Lu}}, \ and\ \bibinfo {author} {\bibfnamefont {W.}~\bibnamefont
  {Ji}},\ }\href@noop {} {\bibfield  {journal} {\bibinfo  {journal} {{P}hys.
  {R}ev. {B}}\ }\textbf {\bibinfo {volume} {99}},\ \bibinfo {pages} {144401}
  (\bibinfo {year} {2019})}\BibitemShut {NoStop}%
\bibitem [{\citenamefont {Sivadas}\ \emph {et~al.}(2018)\citenamefont
  {Sivadas}, \citenamefont {Okamoto}, \citenamefont {Xu}, \citenamefont
  {Fennie},\ and\ \citenamefont {Xiao}}]{Sivadas2018stacking}%
  \BibitemOpen
  \bibfield  {author} {\bibinfo {author} {\bibfnamefont {N.}~\bibnamefont
  {Sivadas}}, \bibinfo {author} {\bibfnamefont {S.}~\bibnamefont {Okamoto}},
  \bibinfo {author} {\bibfnamefont {X.}~\bibnamefont {Xu}}, \bibinfo {author}
  {\bibfnamefont {C.~J.}\ \bibnamefont {Fennie}}, \ and\ \bibinfo {author}
  {\bibfnamefont {D.}~\bibnamefont {Xiao}},\ }\href@noop {} {\bibfield
  {journal} {\bibinfo  {journal} {{N}ano {L}ett.}\ }\textbf {\bibinfo {volume}
  {18}},\ \bibinfo {pages} {7658} (\bibinfo {year} {2018})}\BibitemShut
  {NoStop}%
\bibitem [{\citenamefont {Huang}\ \emph {et~al.}(2018)\citenamefont {Huang},
  \citenamefont {Feng}, \citenamefont {Wu}, \citenamefont {Ahmed},
  \citenamefont {Huang}, \citenamefont {Xiang}, \citenamefont {Deng},\ and\
  \citenamefont {Kan}}]{Chengxi2018Toward}%
  \BibitemOpen
  \bibfield  {author} {\bibinfo {author} {\bibfnamefont {C.}~\bibnamefont
  {Huang}}, \bibinfo {author} {\bibfnamefont {J.}~\bibnamefont {Feng}},
  \bibinfo {author} {\bibfnamefont {F.}~\bibnamefont {Wu}}, \bibinfo {author}
  {\bibfnamefont {D.}~\bibnamefont {Ahmed}}, \bibinfo {author} {\bibfnamefont
  {B.}~\bibnamefont {Huang}}, \bibinfo {author} {\bibfnamefont
  {H.}~\bibnamefont {Xiang}}, \bibinfo {author} {\bibfnamefont
  {K.}~\bibnamefont {Deng}}, \ and\ \bibinfo {author} {\bibfnamefont
  {E.}~\bibnamefont {Kan}},\ }\href@noop {} {\bibfield  {journal} {\bibinfo
  {journal} {{J}. {A}m. {C}hem. {S}oc.}\ }\textbf {\bibinfo {volume} {140}},\
  \bibinfo {pages} {11519} (\bibinfo {year} {2018})}\BibitemShut {NoStop}%
\bibitem [{\citenamefont {Zhuang}\ \emph {et~al.}(2016)\citenamefont {Zhuang},
  \citenamefont {Kent},\ and\ \citenamefont {Hennig}}]{Zhuang2016anisotropy}%
  \BibitemOpen
  \bibfield  {author} {\bibinfo {author} {\bibfnamefont {H.~L.}\ \bibnamefont
  {Zhuang}}, \bibinfo {author} {\bibfnamefont {P.~R.~C.}\ \bibnamefont {Kent}},
  \ and\ \bibinfo {author} {\bibfnamefont {R.~G.}\ \bibnamefont {Hennig}},\
  }\href@noop {} {\bibfield  {journal} {\bibinfo  {journal} {{P}hys. {R}ev.
  {B}}\ }\textbf {\bibinfo {volume} {93}},\ \bibinfo {pages} {134407} (\bibinfo
  {year} {2016})}\BibitemShut {NoStop}%
\bibitem [{\citenamefont {Yang}\ \emph {et~al.}(2020)\citenamefont {Yang},
  \citenamefont {Fan}, \citenamefont {Wang}, \citenamefont {Khomskii},\ and\
  \citenamefont {Wu}}]{yang2020VI3}%
  \BibitemOpen
  \bibfield  {author} {\bibinfo {author} {\bibfnamefont {K.}~\bibnamefont
  {Yang}}, \bibinfo {author} {\bibfnamefont {F.}~\bibnamefont {Fan}}, \bibinfo
  {author} {\bibfnamefont {H.}~\bibnamefont {Wang}}, \bibinfo {author}
  {\bibfnamefont {D.~I.}\ \bibnamefont {Khomskii}}, \ and\ \bibinfo {author}
  {\bibfnamefont {H.}~\bibnamefont {Wu}},\ }\href@noop {} {\bibfield  {journal}
  {\bibinfo  {journal} {{P}hys. {R}ev. {B}}\ }\textbf {\bibinfo {volume}
  {101}},\ \bibinfo {pages} {100402} (\bibinfo {year} {2020})}\BibitemShut
  {NoStop}%
\bibitem [{\citenamefont {Liu}\ \emph {et~al.}(2020)\citenamefont {Liu},
  \citenamefont {Yang}, \citenamefont {Wang},\ and\ \citenamefont
  {Wu}}]{liu2020VBr3}%
  \BibitemOpen
  \bibfield  {author} {\bibinfo {author} {\bibfnamefont {L.}~\bibnamefont
  {Liu}}, \bibinfo {author} {\bibfnamefont {K.}~\bibnamefont {Yang}}, \bibinfo
  {author} {\bibfnamefont {G.}~\bibnamefont {Wang}}, \ and\ \bibinfo {author}
  {\bibfnamefont {H.}~\bibnamefont {Wu}},\ }\href@noop {} {\bibfield  {journal}
  {\bibinfo  {journal} {J. Mater. Chem. C}\ }\textbf {\bibinfo {volume} {8}},\
  \bibinfo {pages} {14782} (\bibinfo {year} {2020})}\BibitemShut {NoStop}%
\bibitem [{\citenamefont {Burch}\ \emph {et~al.}(2018)\citenamefont {Burch},
  \citenamefont {Mandrus},\ and\ \citenamefont {Park}}]{burch2018review}%
  \BibitemOpen
  \bibfield  {author} {\bibinfo {author} {\bibfnamefont {K.~S.}\ \bibnamefont
  {Burch}}, \bibinfo {author} {\bibfnamefont {D.}~\bibnamefont {Mandrus}}, \
  and\ \bibinfo {author} {\bibfnamefont {J.-G.}\ \bibnamefont {Park}},\
  }\href@noop {} {\bibfield  {journal} {\bibinfo  {journal} {Nature}\ }\textbf
  {\bibinfo {volume} {563}},\ \bibinfo {pages} {47} (\bibinfo {year}
  {2018})}\BibitemShut {NoStop}%
\bibitem [{\citenamefont {Cheng}\ and\ \citenamefont
  {Xiang}(2019)}]{Cheng2019review}%
  \BibitemOpen
  \bibfield  {author} {\bibinfo {author} {\bibfnamefont {G.}~\bibnamefont
  {Cheng}}\ and\ \bibinfo {author} {\bibfnamefont {Z.}~\bibnamefont {Xiang}},\
  }\href@noop {} {\bibfield  {journal} {\bibinfo  {journal} {Science}\ }\textbf
  {\bibinfo {volume} {363}},\ \bibinfo {pages} {eaav4450} (\bibinfo {year}
  {2019})}\BibitemShut {NoStop}%
\bibitem [{\citenamefont {Zhao}\ \emph {et~al.}(2018)\citenamefont {Zhao},
  \citenamefont {Huang}, \citenamefont {Peng}, \citenamefont {Huang},
  \citenamefont {Ma},\ and\ \citenamefont {Zhang}}]{Zhao2018MOF2D}%
  \BibitemOpen
  \bibfield  {author} {\bibinfo {author} {\bibfnamefont {M.}~\bibnamefont
  {Zhao}}, \bibinfo {author} {\bibfnamefont {Y.}~\bibnamefont {Huang}},
  \bibinfo {author} {\bibfnamefont {Y.}~\bibnamefont {Peng}}, \bibinfo {author}
  {\bibfnamefont {Z.}~\bibnamefont {Huang}}, \bibinfo {author} {\bibfnamefont
  {Q.}~\bibnamefont {Ma}}, \ and\ \bibinfo {author} {\bibfnamefont
  {H.}~\bibnamefont {Zhang}},\ }\href@noop {} {\bibfield  {journal} {\bibinfo
  {journal} {{C}hem. {S}oc. {R}ev.}\ }\textbf {\bibinfo {volume} {47}},\
  \bibinfo {pages} {6267} (\bibinfo {year} {2018})}\BibitemShut {NoStop}%
\bibitem [{\citenamefont {Luo}\ \emph {et~al.}(2019)\citenamefont {Luo},
  \citenamefont {Ahmad}, \citenamefont {Schug},\ and\ \citenamefont
  {Tsotsalas}}]{Luo2019HierarchicalMOF}%
  \BibitemOpen
  \bibfield  {author} {\bibinfo {author} {\bibfnamefont {Y.}~\bibnamefont
  {Luo}}, \bibinfo {author} {\bibfnamefont {M.}~\bibnamefont {Ahmad}}, \bibinfo
  {author} {\bibfnamefont {A.}~\bibnamefont {Schug}}, \ and\ \bibinfo {author}
  {\bibfnamefont {M.}~\bibnamefont {Tsotsalas}},\ }\href@noop {} {\bibfield
  {journal} {\bibinfo  {journal} {Adv. Mater.}\ }\textbf {\bibinfo {volume}
  {31}},\ \bibinfo {pages} {1901744} (\bibinfo {year} {2019})}\BibitemShut
  {NoStop}%
\bibitem [{\citenamefont {Pedersen}\ \emph {et~al.}(2018)\citenamefont
  {Pedersen}, \citenamefont {Perlepe}, \citenamefont {Aubrey}, \citenamefont
  {Woodruff}, \citenamefont {Reyes-Lillo}, \citenamefont {Reinholdt},
  \citenamefont {Voigt}, \citenamefont {Li}, \citenamefont {Borup},
  \citenamefont {Rouzi{\`e}res}, \citenamefont {Samohvalov}, \citenamefont
  {Wilhelm}, \citenamefont {Rogalev}, \citenamefont {Neaton}, \citenamefont
  {Long},\ and\ \citenamefont {Cl{\'e}rac}}]{pedersen2018formation}%
  \BibitemOpen
  \bibfield  {author} {\bibinfo {author} {\bibfnamefont {K.~S.}\ \bibnamefont
  {Pedersen}}, \bibinfo {author} {\bibfnamefont {P.}~\bibnamefont {Perlepe}},
  \bibinfo {author} {\bibfnamefont {M.~L.}\ \bibnamefont {Aubrey}}, \bibinfo
  {author} {\bibfnamefont {D.~N.}\ \bibnamefont {Woodruff}}, \bibinfo {author}
  {\bibfnamefont {S.~E.}\ \bibnamefont {Reyes-Lillo}}, \bibinfo {author}
  {\bibfnamefont {A.}~\bibnamefont {Reinholdt}}, \bibinfo {author}
  {\bibfnamefont {L.}~\bibnamefont {Voigt}}, \bibinfo {author} {\bibfnamefont
  {Z.}~\bibnamefont {Li}}, \bibinfo {author} {\bibfnamefont {K.}~\bibnamefont
  {Borup}}, \bibinfo {author} {\bibfnamefont {M.}~\bibnamefont
  {Rouzi{\`e}res}}, \bibinfo {author} {\bibfnamefont {D.}~\bibnamefont
  {Samohvalov}}, \bibinfo {author} {\bibfnamefont {F.}~\bibnamefont {Wilhelm}},
  \bibinfo {author} {\bibfnamefont {A.}~\bibnamefont {Rogalev}}, \bibinfo
  {author} {\bibfnamefont {J.~B.}\ \bibnamefont {Neaton}}, \bibinfo {author}
  {\bibfnamefont {J.~R.}\ \bibnamefont {Long}}, \ and\ \bibinfo {author}
  {\bibfnamefont {R.}~\bibnamefont {Cl{\'e}rac}},\ }\href@noop {} {\bibfield
  {journal} {\bibinfo  {journal} {{N}at. {C}hem.}\ }\textbf {\bibinfo {volume}
  {10}},\ \bibinfo {pages} {1056} (\bibinfo {year} {2018})}\BibitemShut
  {NoStop}%
\bibitem [{\citenamefont {Perlepe}\ \emph {et~al.}(2020)\citenamefont
  {Perlepe}, \citenamefont {Oyarzabal}, \citenamefont {Mailman}, \citenamefont
  {Yquel}, \citenamefont {Platunov}, \citenamefont {Dovgaliuk}, \citenamefont
  {Rouzieres}, \citenamefont {Negrier}, \citenamefont {Mondieig}, \citenamefont
  {Suturina}, \citenamefont {Dourges}, \citenamefont {Bonhommeau},
  \citenamefont {Musgrave}, \citenamefont {Pedersen}, \citenamefont
  {Chernyshov}, \citenamefont {Wilhelm}, \citenamefont {Rogalev}, \citenamefont
  {Mathoniere},\ and\ \citenamefont {Clerac}}]{perlepe2020}%
  \BibitemOpen
  \bibfield  {author} {\bibinfo {author} {\bibfnamefont {P.}~\bibnamefont
  {Perlepe}}, \bibinfo {author} {\bibfnamefont {I.}~\bibnamefont {Oyarzabal}},
  \bibinfo {author} {\bibfnamefont {A.}~\bibnamefont {Mailman}}, \bibinfo
  {author} {\bibfnamefont {M.}~\bibnamefont {Yquel}}, \bibinfo {author}
  {\bibfnamefont {M.}~\bibnamefont {Platunov}}, \bibinfo {author}
  {\bibfnamefont {I.}~\bibnamefont {Dovgaliuk}}, \bibinfo {author}
  {\bibfnamefont {M.}~\bibnamefont {Rouzieres}}, \bibinfo {author}
  {\bibfnamefont {P.}~\bibnamefont {Negrier}}, \bibinfo {author} {\bibfnamefont
  {D.}~\bibnamefont {Mondieig}}, \bibinfo {author} {\bibfnamefont {E.~A.}\
  \bibnamefont {Suturina}}, \bibinfo {author} {\bibfnamefont {M.-A.}\
  \bibnamefont {Dourges}}, \bibinfo {author} {\bibfnamefont {S.}~\bibnamefont
  {Bonhommeau}}, \bibinfo {author} {\bibfnamefont {R.~A.}\ \bibnamefont
  {Musgrave}}, \bibinfo {author} {\bibfnamefont {K.~S.}\ \bibnamefont
  {Pedersen}}, \bibinfo {author} {\bibfnamefont {D.}~\bibnamefont
  {Chernyshov}}, \bibinfo {author} {\bibfnamefont {F.}~\bibnamefont {Wilhelm}},
  \bibinfo {author} {\bibfnamefont {A.}~\bibnamefont {Rogalev}}, \bibinfo
  {author} {\bibfnamefont {C.}~\bibnamefont {Mathoniere}}, \ and\ \bibinfo
  {author} {\bibfnamefont {R.}~\bibnamefont {Clerac}},\ }\href@noop {}
  {\bibfield  {journal} {\bibinfo  {journal} {Science}\ }\textbf {\bibinfo
  {volume} {370}},\ \bibinfo {pages} {587} (\bibinfo {year}
  {2020})}\BibitemShut {NoStop}%
\bibitem [{\citenamefont {Kresse}\ and\ \citenamefont
  {Furthm{\"u}ller}(1996)}]{kresse1996}%
  \BibitemOpen
  \bibfield  {author} {\bibinfo {author} {\bibfnamefont {G.}~\bibnamefont
  {Kresse}}\ and\ \bibinfo {author} {\bibfnamefont {J.}~\bibnamefont
  {Furthm{\"u}ller}},\ }\href@noop {} {\bibfield  {journal} {\bibinfo
  {journal} {Phys. Rev. B}\ }\textbf {\bibinfo {volume} {54}},\ \bibinfo
  {pages} {169} (\bibinfo {year} {1996})}\BibitemShut {NoStop}%
\bibitem [{\citenamefont {Perdew}\ \emph {et~al.}(1996)\citenamefont {Perdew},
  \citenamefont {Burke},\ and\ \citenamefont {Ernzerhof}}]{PBE}%
  \BibitemOpen
  \bibfield  {author} {\bibinfo {author} {\bibfnamefont {J.~P.}\ \bibnamefont
  {Perdew}}, \bibinfo {author} {\bibfnamefont {K.}~\bibnamefont {Burke}}, \
  and\ \bibinfo {author} {\bibfnamefont {M.}~\bibnamefont {Ernzerhof}},\
  }\href@noop {} {\bibfield  {journal} {\bibinfo  {journal} {Phys. Rev. Lett.}\
  }\textbf {\bibinfo {volume} {77}},\ \bibinfo {pages} {3865} (\bibinfo {year}
  {1996})}\BibitemShut {NoStop}%
\bibitem [{\citenamefont {Anisimov}\ \emph {et~al.}(1991)\citenamefont
  {Anisimov}, \citenamefont {Zaanen},\ and\ \citenamefont {Andersen}}]{U}%
  \BibitemOpen
  \bibfield  {author} {\bibinfo {author} {\bibfnamefont {V.~I.}\ \bibnamefont
  {Anisimov}}, \bibinfo {author} {\bibfnamefont {J.}~\bibnamefont {Zaanen}}, \
  and\ \bibinfo {author} {\bibfnamefont {O.~K.}\ \bibnamefont {Andersen}},\
  }\href@noop {} {\bibfield  {journal} {\bibinfo  {journal} {Phys. Rev. B}\
  }\textbf {\bibinfo {volume} {44}},\ \bibinfo {pages} {943} (\bibinfo {year}
  {1991})}\BibitemShut {NoStop}%
\bibitem [{\citenamefont {Grimme}(2006)}]{VdW-D2}%
  \BibitemOpen
  \bibfield  {author} {\bibinfo {author} {\bibfnamefont {S.}~\bibnamefont
  {Grimme}},\ }\href@noop {} {\bibfield  {journal} {\bibinfo  {journal} {J.
  Comput. Chem.}\ }\textbf {\bibinfo {volume} {27}},\ \bibinfo {pages} {1787}
  (\bibinfo {year} {2006})}\BibitemShut {NoStop}%
\bibitem [{\citenamefont {McGuire}\ \emph {et~al.}(2015)\citenamefont
  {McGuire}, \citenamefont {Dixit}, \citenamefont {Cooper},\ and\ \citenamefont
  {Sales}}]{mcguire2015coupling}%
  \BibitemOpen
  \bibfield  {author} {\bibinfo {author} {\bibfnamefont {M.~A.}\ \bibnamefont
  {McGuire}}, \bibinfo {author} {\bibfnamefont {H.}~\bibnamefont {Dixit}},
  \bibinfo {author} {\bibfnamefont {V.~R.}\ \bibnamefont {Cooper}}, \ and\
  \bibinfo {author} {\bibfnamefont {B.~C.}\ \bibnamefont {Sales}},\ }\href@noop
  {} {\bibfield  {journal} {\bibinfo  {journal} {Chem. Mater.}\ }\textbf
  {\bibinfo {volume} {27}},\ \bibinfo {pages} {612} (\bibinfo {year}
  {2015})}\BibitemShut {NoStop}%
\bibitem [{\citenamefont {Wadt}\ and\ \citenamefont
  {Goddard}(1975)}]{1975Electronic}%
  \BibitemOpen
  \bibfield  {author} {\bibinfo {author} {\bibfnamefont {W.~R.}\ \bibnamefont
  {Wadt}}\ and\ \bibinfo {author} {\bibfnamefont {W.~A.}\ \bibnamefont
  {Goddard}},\ }\href@noop {} {\bibfield  {journal} {\bibinfo  {journal} {{J}.
  {A}m. {C}hem. {S}oc.}\ }\textbf {\bibinfo {volume} {97}},\ \bibinfo {pages}
  {2034} (\bibinfo {year} {1975})}\BibitemShut {NoStop}%
\bibitem [{\citenamefont {Jiang}\ \emph {et~al.}(2018)\citenamefont {Jiang},
  \citenamefont {Li}, \citenamefont {Wang}, \citenamefont {Mak},\ and\
  \citenamefont {Shan}}]{jiang2018controlling}%
  \BibitemOpen
  \bibfield  {author} {\bibinfo {author} {\bibfnamefont {S.}~\bibnamefont
  {Jiang}}, \bibinfo {author} {\bibfnamefont {L.}~\bibnamefont {Li}}, \bibinfo
  {author} {\bibfnamefont {Z.}~\bibnamefont {Wang}}, \bibinfo {author}
  {\bibfnamefont {K.~F.}\ \bibnamefont {Mak}}, \ and\ \bibinfo {author}
  {\bibfnamefont {J.}~\bibnamefont {Shan}},\ }\href@noop {} {\bibfield
  {journal} {\bibinfo  {journal} {Nat. Nanotechnol.}\ }\textbf {\bibinfo
  {volume} {13}},\ \bibinfo {pages} {549} (\bibinfo {year} {2018})}\BibitemShut
  {NoStop}%
\bibitem [{\citenamefont {Cao}\ \emph {et~al.}(2015)\citenamefont {Cao},
  \citenamefont {Li},\ and\ \citenamefont {Louie}}]{Cao2015doping}%
  \BibitemOpen
  \bibfield  {author} {\bibinfo {author} {\bibfnamefont {T.}~\bibnamefont
  {Cao}}, \bibinfo {author} {\bibfnamefont {Z.}~\bibnamefont {Li}}, \ and\
  \bibinfo {author} {\bibfnamefont {S.~G.}\ \bibnamefont {Louie}},\ }\href@noop
  {} {\bibfield  {journal} {\bibinfo  {journal} {Phys. Rev. Lett.}\ }\textbf
  {\bibinfo {volume} {114}},\ \bibinfo {pages} {236602} (\bibinfo {year}
  {2015})}\BibitemShut {NoStop}%
\end{thebibliography}%

\newpage
\clearpage

\begin{appendix}
\setcounter{figure}{0}
\setcounter{table}{0}
\renewcommand{\thefigure}{S\arabic{figure}}
\renewcommand{\thetable}{S\arabic{table}}
\renewcommand{\tablename}{Table}
\renewcommand{\figurename}{Fig.}
%\title{Supplementary Material: VI$_{3}$: a 2D Ising ferromagnet}
%\maketitle
\section{Supporting Information for ``2D hybrid CrCl$_2$(N$_2$C$_4$H$_4$)$_2$ with tunable ferromagnetic half-metallicity''}

\begin{table}[H]
%\scriptsize
%\footnotesize
%\tiny
  \caption{Relative total energy $\delta$ E (meV/f.u.) and lattice constants ($a$, $b$, and $c$ in unit of \AA) for bulk structures $\alpha$, $\beta$, $\gamma$, and $\delta$ of CrCl$_2$(pyrazine)$_2$ with full structural optimization, see also Fig. 5 in the main text. The optimized lattice constants of the most stable $\alpha$ structure agree well with the experimental ones of $a$ = 6.90 \AA, $b$ =6.97 \AA and $c$ = 10.83 \AA (Ref. 37).}
  \label{tb1}
\begin{tabular}{l@{\hskip8mm}c@{\hskip8mm}c@{\hskip8mm}c@{\hskip8mm}c@{\hskip8mm}}
\hline\hline
Bulk &   $\Delta$E                 & $a$ & $b$  & $c$    \\ \hline
$\alpha$&   0                      & 6.87 & 6.88     & 10.36   \\
$\beta$&    248                    & 6.93  &  6.94   & 10.45 \\
$\gamma$&   $\rightarrow$ $\alpha$ &  &    &            \\
$\delta$&   $\rightarrow$ $\alpha$ &  &    &        \\
\hline\hline
 \end{tabular}
\end{table}

\begin{figure}[H]
\includegraphics[width=7cm]{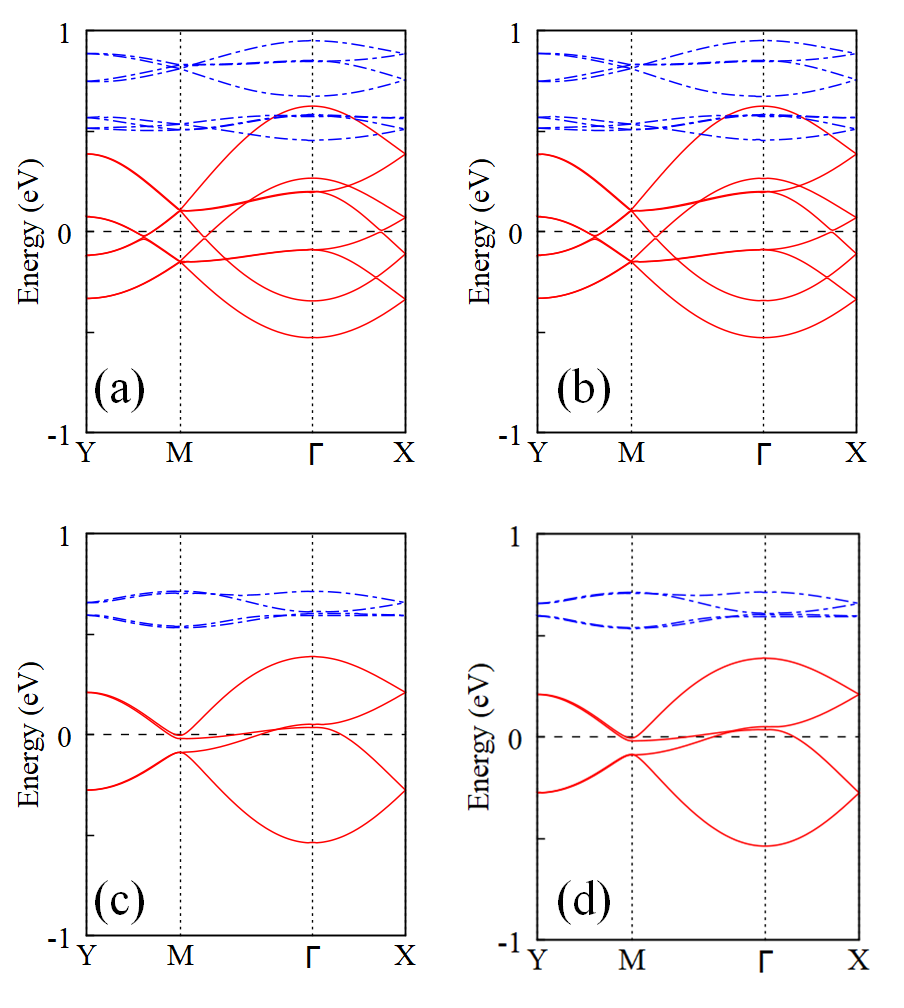}
 \caption{Band structure of bulk CrCl$_2$(pyrazine)$_2$ calculated by (a) GGA +U and (b) GGA+U+SOC. Band  structure of monolayer CrCl$_2$(pyrazine)$_2$ calculated by (c) GGA +U and (d) GGA+U+SOC. The blue (red)  lines stand for the up (down) spin. The Fermi level is set at zero energy.
}
\end{figure}

\begin{figure}[H]
\includegraphics[width=7cm]{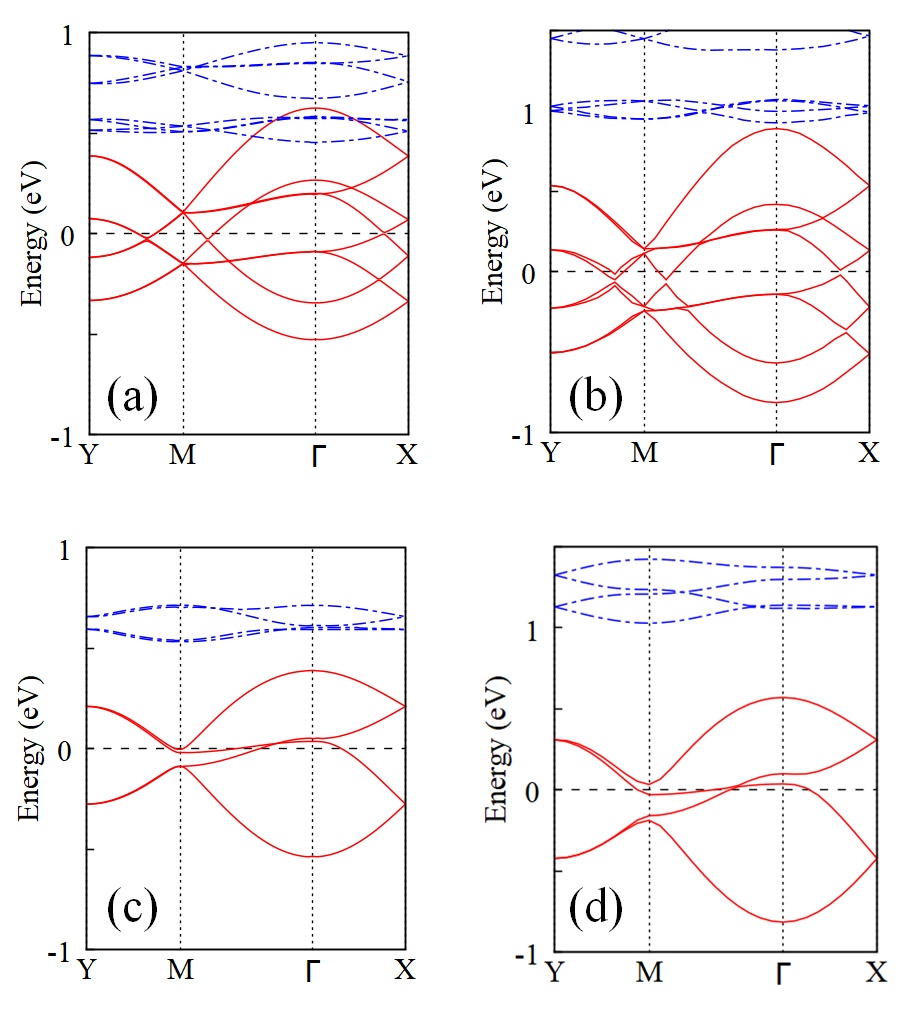}

\caption{ Band structure of bulk CrCl$_2$(pyrazine)$_2$ calculated by (a) GGA + U and (b) HSE06. Band structure of monolayer CrCl$_2$(pyrazine)$_2$ calculated by (c) GGA + U and (d) HSE06. The blue (red) lines stand for the up (down) spin. The Fermi level is set at zero energy.
 }
   \label{hse}
\end{figure}

\end{appendix}

\end{document}